\documentclass[journal,onecolumn,11pt]{IEEEtran}

\usepackage{graphics}
\usepackage{amssymb}
\usepackage{amsmath}
\usepackage{amsfonts}
\usepackage{physics}
\usepackage{graphicx}
\usepackage{fancybox}
\usepackage{accents}
\usepackage{times}
\usepackage{color}

\usepackage{algorithm}
\usepackage[noend]{algpseudocode}
\usepackage{array}
\usepackage{textcomp}
\usepackage{stfloats}
\usepackage{float}
\usepackage{verbatim}
\usepackage{microtype}

\usepackage[nolist, nohyperlinks]{acronym}
\usepackage{graphbox}
\usepackage[inkscapeformat=png]{svg}

\usepackage{setspace} 
\usepackage{caption}
\usepackage{subcaption}

\usepackage{comment}
\usepackage{cite}
\usepackage{soul}
\usepackage{epstopdf}
\usepackage{xfrac}
\usepackage{pifont}

\usepackage{cleveref}
\usepackage{csquotes}
\usepackage{booktabs}
\usepackage{multirow}
\usepackage{tabularx,ragged2e}
\newcolumntype{Y}{>{\centering\arraybackslash}X}
\usepackage{footmisc}

\usepackage{varwidth}
\usepackage{tikz}
\usepackage{pgf}
\usepackage{pgfplots}
\usepackage{pgfplotstable}
\usetikzlibrary{positioning,backgrounds,fit,calc,patterns,arrows.meta}
\pgfplotsset{compat=1.3}
\usetikzlibrary{shapes.geometric}
\usetikzlibrary{shapes.arrows}
\usetikzlibrary{arrows,decorations.markings}
\usepackage[skins]{tcolorbox}

\begin{acronym}
\acro{stft}[STFT]{short-time Fourier transform}
\acro{istft}[iSTFT]{inverse short-time Fourier transform}
\acro{cqt}[CQT]{Constant-Q Transform}
\acro{dnn}[DNN]{deep neural network}
\acro{pesq}[PESQ]{Perceptual Evaluation of Speech Quality}
\acro{polqa}[POLQA]{perceptual objectve listening quality analysis}
\acro{wpe}[WPE]{weighted prediction error}
\acro{psd}[PSD]{power spectral density}
\acro{rir}[RIR]{room impulse response}
\acro{snr}[SNR]{signal-to-noise ratio}
\acro{lstm}[LSTM]{long short-term memory}
\acro{polqa}[POLQA]{Perceptual Objectve Listening Quality Analysis}
\acro{sdr}[SDR]{signal-to-distortion ratio}
\acro{estoi}[ESTOI]{Extended Short-Term Objective Intelligibility}
\acro{elr}[ELR]{early-to-late reverberation ratio}
\acro{tcn}[TCN]{temporal convolutional network}
\acro{rls}[RLS]{recursive least squares}
\acro{asr}[ASR]{automatic speech recognition}
\acro{ha}[HA]{hearing aid}
\acro{ci}[CI]{cochlear implant}
\acro{mac}[MAC]{multiply-and-accumulate}
\acro{mmse}[MMSE]{minimum mean square error}
\acro{vae}[VAE]{variational auto-encoder}
\acro{gan}[GAN]{generative adversarial network}
\acro{tf}[T-F]{time-frequency}
\acro{sde}[SDE]{stochastic differential equation}
\acro{ode}[ODE]{ordinary differential equation}
\acro{drr}[DRR]{direct to reverberant ratio}
\acro{lsd}[LSD]{log spectral distance}
\acro{sisdr}[SI-SDR]{scale-invariant signal to distortion ratio}
\acro{mos}[MOS]{mean opinion score}
\acro{map}[MAP]{maximum a posteriori}
\acro{rtf}[RTF]{real-time factor}
\acro{ddpm}[DDPM]{denoising diffusion probabilistic model}
\acro{map}[MAP]{maximum a posteriori}
\acro{sgmse}[SGMSE]{Score-based generative models for speech enhancement}
\end{acronym}

\newcommand{\x}[0]{\mathbf{x}}
\newcommand{\y}[0]{\mathbf{y}}
\newcommand{\n}[0]{\mathbf{n}}

\newcommand{\cond}[0]{\mathbf{c}}

\newcommand*\diff{\mathop{}\!\mathrm{d}}

\newcommand{\state}[0]{\mathbf{x}_{\tau}}
\newcommand{\wien}[0]{\mathbf{w}_\tau}
\newcommand{\clean}[0]{\mathbf{x}_0}
\newcommand{\init}[0]{\mathbf{x}_T}

\newcommand{\sco}[0]{\nabla_{\state} \log p_\tau(\state)}
\newcommand{\postsco}[0]{\nabla_{\state} \log p_\tau(\state  | \y)}
\newcommand{\likelihood}[0]{\nabla_{\state} \log p_\tau(\y | \state)}
\newcommand{\scomodel}[0]{\mathbf{s}_\theta(\state, \tau)}
\newcommand{\kernel}[0]{q_\tau( \state | \clean)}

\makeatletter
\newcommand{\subalign}[1]{%
  \vcenter{%
    \Let@ \restore@math@cr \default@tag
    \baselineskip\fontdimen10 \scriptfont\tw@
    \advance\baselineskip\fontdimen12 \scriptfont\tw@
    \lineskip\thr@@\fontdimen8 \scriptfont\thr@@
    \lineskiplimit\lineskip
    \ialign{\hfil$\m@th\scriptstyle##$&$\m@th\scriptstyle{}##$\hfil\crcr
      #1\crcr
    }%
  }%
}

\begin{document}
\title{Diffusion Models for Audio Restoration \\
{\normalsize{Invited paper for the SPM Special entitled ``Model-based and Data-Driven Audio Signal Processing".}}}

\author{Jean-Marie Lemercier$^\dagger$, Julius Richter$^\dagger$, Simon Welker$^{\dagger}$, Eloi Moliner$^\circ$, Vesa Välimäki$^\circ$, Timo Gerkmann$^\dagger$ \\ %
$^\dagger$\normalsize{Signal Processing (SP), Department of Informatics, Universität Hamburg, Germany}\\
$^\circ$\normalsize{Acoustics Lab, Department of Information and Communications Engineering, Aalto University, Espoo, Finland}}

\maketitle

\nocite{godsill_digital_1998, gerkmann2018book_chapter, wang2018supervised, MurphyBook2, goodfellow2014generative, kingma2014auto, ho2020denoising}

\vspace{-2cm}
\begin{figure}[h]
    \centering
    \input{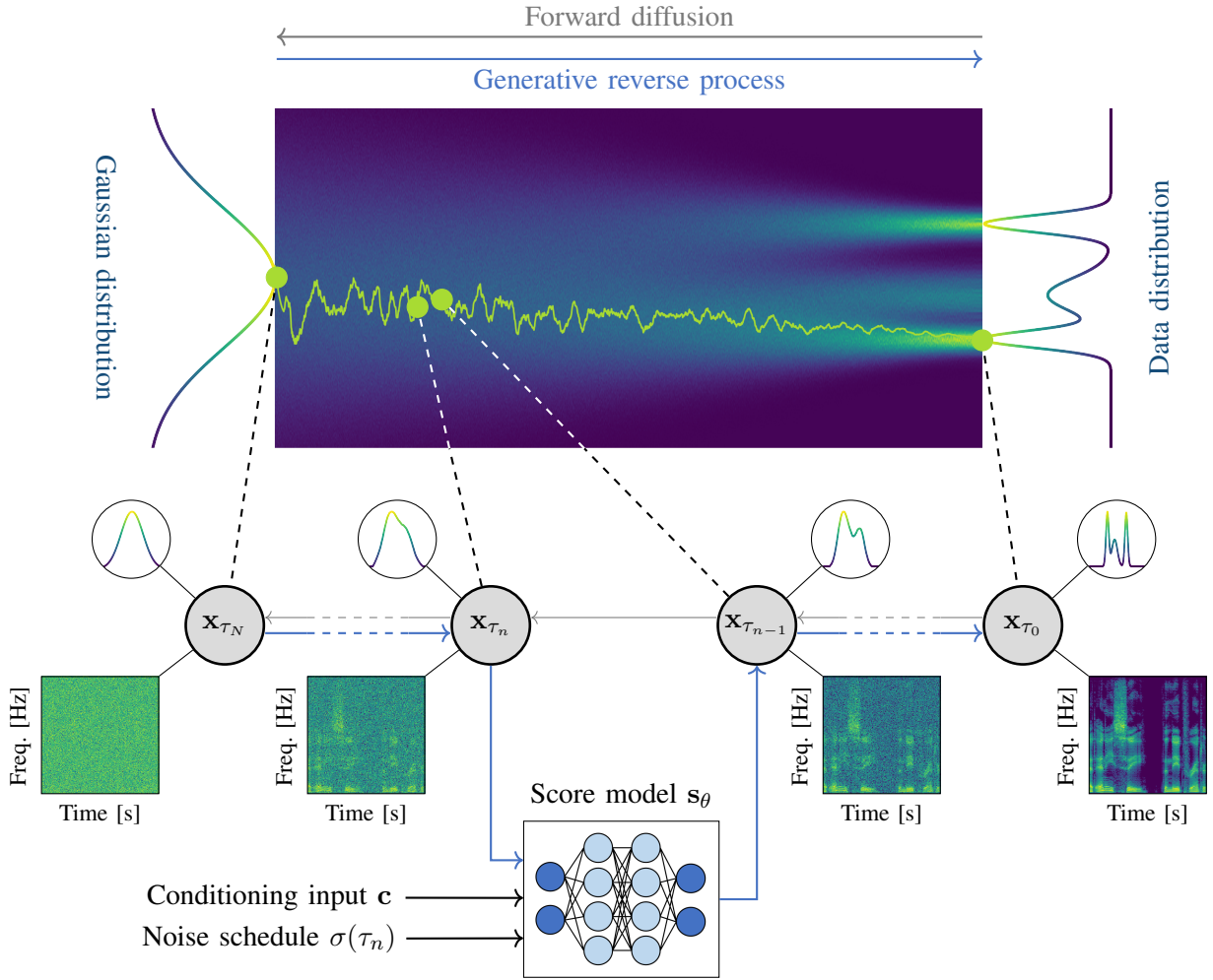}
    \vspace{0.1cm}
    \caption{A continuous-time diffusion model~\cite{song2021sde} transforms (left) a Gaussian distribution to (right) an intractable data distribution through a stochastic process $\{ \state \}_{\tau \in [0, T]}$ with marginal distributions $\{ p_\tau(\state) \}_{\tau \in [0, T]}$. During training, the forward diffusion is simulated by adding Gaussian noise and rescaling the data, and a score model $\mathbf{s}_\theta$ with parameters $\theta$ learns the score function $\sco$. During the generative reverse process, the process time $\tau$ is discretized to steps $\{ \tau_0, \dots, \tau_N \}$ and followed in reverse from $\tau_N = T$ to $\tau_0 = 0$. (Bottom) The next state $\x_{\tau_{n-1}}$ is obtained based on the previous state $\x_{\tau_{n}}$ using an estimate given by the score model. The score model is conditioned by the noise scale at the current time step, $\sigma(\tau_{n})$, and optional conditioning $\cond$ to guide the generation such as, e.g., a text description.
    }
    \label{fig:diffusion_intro}
\end{figure}

\label{secrel}

\clearpage

With the development of audio playback devices and fast data transmission, the demand for high sound quality is rising for both entertainment and communications. In this quest for better sound quality, challenges emerge from distortions and interferences originating at the recording side or caused by an imperfect transmission pipeline. To address this problem, audio restoration methods aim to recover clean sound signals from the corrupted input data. We present here audio restoration algorithms based on diffusion models, with a focus on speech enhancement and music restoration tasks. 

Traditional approaches, often grounded in handcrafted rules and statistical heuristics, have shaped our understanding of audio signals. In the past decades, there has been a notable shift towards data-driven methods that exploit the modeling capabilities of \acp{dnn}. Deep generative models, and among them diffusion models, have emerged as powerful techniques for learning complex data distributions. However, relying solely on \ac{dnn}-based learning approaches carries the risk of reducing interpretability, particularly when employing end-to-end models. Nonetheless, data-driven approaches allow more flexibility in comparison to statistical model-based frameworks, whose performance depends on distributional and statistical assumptions that can be difficult to guarantee. Here, we aim to show that diffusion models can combine the best of both worlds and offer the opportunity to design audio restoration algorithms with a good degree of interpretability and a remarkable performance in terms of sound quality.

In this article, we review the use of diffusion models for audio restoration. We explain the diffusion formalism and its application to the conditional generation of clean audio signals. We believe that diffusion models open an exciting field of research with the potential to spawn new audio restoration algorithms that are natural-sounding and remain robust in difficult acoustic situations.

\section*{Introduction}
\label{sec:introduction}
Traditional audio restoration methods exploit statistical properties of audio signals, such as auto-regressive modeling for click removal \cite{godsill_digital_1998} or probabilistic modeling for speech enhancement and separation \cite{gerkmann2018book_chapter}, by using various representations like time-domain waveforms, spectrograms, or cepstra. Although they are robust to many scenarios, such methods struggle with highly non-stationary sources or interferences that appear in real-life scenarios. In the past decade, audio signal processing algorithms have benefited greatly from the introduction of data-driven approaches based on \acp{dnn}~\cite{wang2018supervised}. Among these methods, a broad class leverages \emph{predictive models} that learn to map a given input to a desired output. Note that the term \emph{predictive models} covers both classification and regression tasks, unlike \textit{discriminative models} \cite{MurphyBook2}. In a typical supervised setting, a predictive model is trained on a labeled dataset to minimize a certain point-wise loss function between the processed input and the clean target. Following the principle of empirical risk minimization, the goal of predictive modeling is to find a model with minimal average error over the training data, where the generalization ability of the model is usually assessed on a validation set of unseen data. By employing ever-larger models and datasets---a current trend in deep learning---strong generalization can be achieved. However, many purely data-driven approaches are considered black boxes and remain largely unexplainable and non-interpretable. Moreover, these models typically produce deterministic outputs, disregarding the inherent uncertainty in their results.

\emph{Generative models} follow a different learning paradigm, namely estimating and sampling from an unknown data distribution. This can be used to infer a measure of uncertainty for their predictions and to allow the generation of multiple valid estimates instead of a single best estimate as in predictive approaches \cite{MurphyBook2}. Furthermore, incorporating prior knowledge into generative models can guide the learning process and enforce desired properties about the learned distribution. \Acp{gan}~\cite{goodfellow2014generative} and variational autoencoders~\cite{kingma2014auto} have been instrumental in the early developments of generative models. Subsequent to these approaches, \textit{diffusion models} \cite{ho2020denoising, song2021sde} have emerged as a distinct class of deep generative models that boast an impressive ability to learn complex data distributions such as that of natural images \cite{ho2020denoising, song2021sde}, music \cite{moliner_solving_2022}, and speech \cite{chen2021wavegrad}. Diffusion models generate data samples through iterative transformations, transitioning from a tractable prior distribution (e.g., Gaussian) to a target data distribution, as visualized in Figure \ref{fig:diffusion_intro}. This iterative generation scheme is formalized as a stochastic process and is parameterized with a \ac{dnn} that is trained to address a Gaussian denoising task.

From a practical point of view, diffusion models have become popular because they can generate high-quality samples while being simpler to train than \acp{gan}. Moreover, combining data-driven machine learning techniques with mathematical concepts, such as stochastic processes, opens up possibilities for modeling conditional data distributions and integrating Bayesian inference tools. In audio processing, this has spawned new types of algorithms that adopt diffusion models for restoration tasks such as speech enhancement~\cite{lu2022conditional, richter2023speech} or music restoration \cite{moliner_solving_2022}. Here, we present a comprehensive overview and categorization of novel techniques for solving audio restoration problems using diffusion models in a data-driven, model-based fashion. 

In the following, we first look at the basics of diffusion models and show how they can be used for model-based processing. We then examine conditional generation with diffusion models for audio restoration tasks, distinguishing between three different conditioning techniques. In particular, we look at diffusion models for audio inverse problems with a known degradation operator and its extension to blind inverse problems when the degradation operator is unknown. We conclude by discussing the practical requirements of diffusion models for audio restoration tasks, examining sampling speed and robustness to adverse conditions.

\section*{Basics of Diffusion Models}%
\label{sec:basics}
With the development of \acp{dnn} and the increase in computational power, deep generative modeling has become one of the leading directions in machine learning with a variety of applications. Deep generative models aim to design a generation process for data that resembles real-world examples, e.g., natural speech produced by a human speaker. This involves modeling the probability distribution of highly structured and complex data such that learning and sampling are computationally tractable. One way to realize generative modeling is based on the assumption that the data is generated by some random process involving unobserved variables. Such \emph{hidden variable models} map samples from a tractable distribution, such as the Gaussian distribution, to samples that are likely to represent target data points. From this perspective of hidden variable models, we discuss diffusion models as a distinct class of deep generative models whose hidden variables are parameterized via a stochastic process.

Diffusion models break down the problem of generating high-dimensional complex data into a series of easier \textit{denoising} tasks. Training such a denoising model first requires defining a \emph{forward diffusion process}, which gradually adds noise to the data points of a dataset. This corruption process progressively turns the data distribution into a Gaussian distribution, as shown in Figure~\ref{fig:diffusion_intro} from right to left (gray arrows). In turn, data generation is accomplished by reversing the corruption process. First, a random sample is drawn from a Gaussian distribution, and then the model iteratively removes noise from this initial point, ultimately yielding a sample from the data distribution. This \emph{reverse diffusion process} is illustrated in Figure~\ref{fig:diffusion_intro} from left to right (blue arrows). 

Formally, the forward diffusion process can be represented by the Markov chain $\mathbf x_0 \rightarrow \mathbf x_{1} \rightarrow \dots \rightarrow \mathbf x_{N}$ with $\mathbf x_0 \in \mathbb{R}^d$ sampled from the data distribution and fixed Gaussian transition probabilities $q(\mathbf x_n | \mathbf x_{n-1})$. The resulting directed graphical model is depicted with gray circles in Figure~\ref{fig:diffusion_intro}. The generative model is then described by a Markov chain in reverse order $\mathbf x_N \rightarrow \mathbf x_{N-1} \rightarrow \dots \rightarrow \mathbf x_{0}$ with $\mathbf x_N$ sampled from the prior distribution $\mathcal N(\mathbf 0, \mathbf I)$. To accomplish the generation task, a \ac{dnn} is trained to denoise the sample $\x_{n}$. Specifically, it learns to approximate the transition probabilities of the reverse Markov chain $p(\x_{n-1} | \x_{n})$ \cite{ho2020denoising}.

This discrete-time Markov chain formulation of diffusion models can be generalized to continuous-time stochastic processes by letting the number of steps $N$ grow infinitely, conversely making the distance between steps infinitely small. This facilitates the design of novel diffusion processes and allows the use of more flexible sampling schemes~\cite{song2021sde}. Specifically, the corresponding forward diffusion process is defined as a stochastic process $\{ \state \}_{ \tau \in [0, T] }$, i.e., a collection of random variables indexed by a continuous process time $\tau \in [0, T]$ \cite{Oksendal2000SDE}. The process time $\tau$ in stochastic processes intuitively corresponds to the index $n$ in  Markov chains. It is important to note that the process time $\tau$ is completely unrelated to the time dimension of the audio signal. A single random realization of the stochastic process  $\{ \state \}_{ \tau \in [0, T] }$ is depicted by the green trajectory traversing Figure~\ref{fig:diffusion_intro}. The conditional distribution characterizing the forward diffusion model is the \textit{transition kernel} $\kernel$ instinctively related to the probability $q(\x_n | \x_0) := {\prod_{i = 1}^n} q(\x_i | \x_{i-1})$ in Markov chains. The transition kernel $\kernel$ can be computed by solving a \ac{sde}, which is a differential equation where some of the coefficients are random~\cite{Oksendal2000SDE}. Specifically, we define the so-called \textit{forward SDE} as
\begin{equation} \label{eq:forward-sde}
    \diff{\state} = \mathbf f(\mathbf \state, \tau) \diff{\tau} + g(\tau) \diff{\wien} \,,
\end{equation}
where the function $\mathbf f: \mathbb R^d \times \mathbb R \rightarrow \mathbb R^d$ is referred to as the \emph{drift coefficient} and relates to the deterministic part of the \ac{sde}. The function $g: \mathbb R \rightarrow \mathbb R$ is called the \emph{diffusion coefficient} and controls the amount of randomness in the SDE. More precisely, the diffusion coefficient $g(\tau)$ scales the noise injected by the stochastic process $\wien$. In most cases, $\wien$ is chosen to be a Wiener process, which is a stochastic process with independent and normally distributed increments, i.e., 
$\mathbf{w}_{\tau + \diff{\tau}} - \wien 
\sim \mathcal{N}(\mathbf{0}, \diff{\tau} \, \mathbf{I})$ \cite{Oksendal2000SDE}.
If the drift coefficient $\mathbf f$ is an affine function of $\state$ and the diffusion coefficient $g$ is independent of $\state$, then the transition kernel has a simple Gaussian form 
\begin{equation} \label{eq:transition_kernel}
    \kernel = \mathcal{N}(
    \boldsymbol\mu(\clean, \tau), \sigma(\tau)^2 \mathbf I)\,,
\end{equation}
where the mean $\boldsymbol\mu(\clean, \tau)$ and standard deviation $\sigma(\tau)$ are obtained by analytically solving the SDE and computing the first and second moments of the solution \cite{Oksendal2000SDE}.

The reverse diffusion process, i.e. the generation process, is also a stochastic process $\{ \state \}_{ \tau \in [0, T] }$ parameterized by the process time $\tau \in [0, T]$ flowing in the reverse direction, with $\mathbf x_T \sim \mathcal{N}(\mathbf{0}, \sigma(T)^2 \boldsymbol{I})$. Reversing the process time axis in \eqref{eq:forward-sde} results in another SDE called the \emph{reverse SDE} whose marginal distributions match those of the forward SDE \cite{song2021sde}. Therefore, denoising the sample $\state$ boils down to solving the reverse SDE
\begin{equation}\label{eq:reverse-sde}
\diff{\state} =
        \left[
            \mathbf f(\mathbf \state, \tau) - g(\tau)^2 \sco
        \right]
        \diff{\tau}
        + g(\tau) \diff{\bar{\mathbf w}_\tau} \,,
\end{equation}
where $\diff{\tau} < 0$ as the process axis is traveled in the reverse direction. The stochastic process $\bar{\mathbf{w}}_\tau$ is another Wiener process associated to this reverse process axis, i.e., $\bar{\mathbf{w}}_{\tau + \diff{\tau}} - \bar{\mathbf{w}}_\tau \sim \mathcal{N}(\mathbf{0}, -\diff{\tau} \, \mathbf{I})$. The quantity $\sco$ (with $\nabla_{\state}$ representing the gradient operator with respect to $\state$) is called \textit{score function} and is a vector field informative about the variations of the process state's logarithmic probability density. The score function $\sco$ is generally intractable and we need to approximate it with a \ac{dnn} called \emph{score model} $\scomodel$ with parameters $\theta$.
Vincent et al.~\cite{vincent2011connection} have shown that the score model $\scomodel$ can be optimized using \emph{denoising score matching}, i.e., matching the score of the Gaussian transition kernel $\kernel$ instead of the score of the unknown probability $p_\tau(\state)$. The score of the transition kernel $\kernel$ can be obtained from~\eqref{eq:transition_kernel} as
\begin{equation} \label{eq:score}
    \nabla_{\state} \log \kernel = - \frac{\state - \boldsymbol{\mu}(\clean, \tau) }{\sigma(\tau)^2} \,.
\end{equation}
The score model $\mathbf{s}_\theta$ is therefore trained using the denoising score-matching objective \cite{vincent2011connection}
\begin{equation} \label{eq:training}
\mathbb{E}_{\subalign{&\clean \sim p(\clean) \\ &\tau \sim \mathcal{U}(0, T) \\ &\state \sim \kernel}}
\left[
 \lambda(\tau) 
 \left\lVert
 \scomodel + \frac{\state - \boldsymbol{\mu}(\clean, \tau) }{\sigma(\tau)^2} 
 \right\rVert_2^2
 \right],
\end{equation}
where a data point $\clean$ is first sampled from the training set. Then, a process time $\tau$ is sampled uniformly between $0$ and $T$, and the diffusion state $\state$ is obtained by sampling from the transition kernel \eqref{eq:transition_kernel}. Here $\lambda(\tau)$ is a time-dependent scaling factor, chosen empirically to stabilize training \cite{song2021sde}.

\begin{figure*}[t]
\input{figures/figure_sde_ode_3}
\end{figure*}

Once the score model $\mathbf s_\theta$ has been trained, it allows the generation of new samples from the learned data distribution by solving the reverse \ac{sde} \eqref{eq:reverse-sde}. In practice, this is done by first discretizing the process time axis into $N$ steps $\{ \tau_N, \tau_{N-1}, \dots, \tau_0 \} $ with a step size $\Delta \tau_n := \tau_{n-1} - \tau_n < 0$, often chosen uniformly. Then an initial condition $\mathbf{x}_{\tau_N}$ is sampled and the reverse SDE \eqref{eq:reverse-sde} is integrated between $\tau_N=T$ and $\tau_0=0$ using a numerical approximation method called \textit{SDE solver} \cite{numericalrecipes}. A differential equation solver approximates the trajectory between successive steps $\mathbf{x}_{\tau_n}$ and $\mathbf{x}_{\tau_{n-1}}$ as a piecewise polynomial function (linear if first-order solver, quadratic if second-order, etc.), whose coefficients depend on the terms in~\eqref{eq:reverse-sde}. An SDE solver, in particular, considers two polynomial functions (with potentially distinct degrees) to model the deterministic and stochastic terms, respectively. For instance, the widespread Euler-Maruyama method is an SDE solver with first-order polynomial approximation for both the deterministic and stochastic components \cite{numericalrecipes}. The generation process is summarized in the algorithm in Figure~\ref{fig:stochastic_vs_deterministic_sampling}.

Deterministic sampling can also be used in place of stochastic sampling by deactivating the randomness source, i.e., removing the Wiener process $\bar{\mathbf{w}}_\tau$ in \eqref{eq:reverse-sde} and scaling the diffusion coefficient $g$. This turns the reverse SDE into a so-called \textit{probability flow \ac{ode}} \cite{song2021sde}. A comparison between stochastic and deterministic sampling is presented in Figure~\ref{fig:stochastic_vs_deterministic_sampling}. We display three realizations of the stochastic and deterministic sampler. Note that because of the stochastic noise injected at each step, two stochastic sampler trajectories starting at the same initial state may end up reaching two different modes of the target data distribution, whereas the corresponding mean trajectory systematically reaches the same mode. This suggests that a stochastic sampler could be used to obtain more diverse samples and also to improve mode coverage, i.e., better represent the modes of the data distribution $p(\clean)$ regardless of the initial state.

\begin{figure}
    \centering
    \tikzstyle{mycircle} = [circle, draw, fill=white, inner sep=0pt, minimum size=30pt]

\tikzstyle{mysquare} = [rectangle, draw, fill=white, inner sep=0pt, minimum size=30pt]
\tikzstyle{myrectangle_waveform} = [circle, draw, fill=white, inner sep=0pt, minimum size=30pt]

\tikzstyle{myrectangle3} = [rectangle, draw, fill=white, inner sep=0pt, minimum width=45pt, minimum height=30pt, align=center]
\tikzstyle{mybranch} = [circle, draw, fill=black, inner sep=0pt, minimum size=3pt]

\tikzstyle{sum} = [
  circle,
  draw,
  minimum size=12pt,
  append after command={
    \pgfextra{
      \draw (\tikzlastnode.north) -- (\tikzlastnode.south);
      \draw (\tikzlastnode.west) -- (\tikzlastnode.east);
    }
  },
]
\tikzstyle{sumcolor} = [
  circle,
  draw,
  figdarkergreen,
  minimum size=12pt,
  append after command={
    \pgfextra{
      \draw[figdarkergreen] (\tikzlastnode.north) -- (\tikzlastnode.south);
      \draw[figdarkergreen] (\tikzlastnode.west) -- (\tikzlastnode.east);
    }
  },
]

\newcommand{\scale}{0.82}
\newcommand{\vy}{0.cm}
\newcommand{\sepx}{0.75cm}
\newcommand{\sepxx}{0.4cm}
\newcommand{\sepy}{0.6cm}
\newcommand{\sepyy}{0.2cm}
\newcommand{\sepxxx}{0.cm}

\newcommand{\sepnodex}{0.3cm}
\newcommand{\sepnodey}{0.3cm}

\definecolor{figblue}{HTML}{154c79}
\definecolor{figdarkergreen}{HTML}{000000}

\hspace{-0.3cm}
\begin{minipage}{0.45\textwidth}
    
\centering
\begin{center}
    \textbf{Waveform Diffusion}
\end{center}
\scalebox{\scale}{
\begin{tikzpicture}

\coordinate (pos_x) at (1.8, 0);

\node (input) at (pos_x) {$\clean$};
\node[myrectangle_waveform, left=\sepnodex of input, fill overzoom image={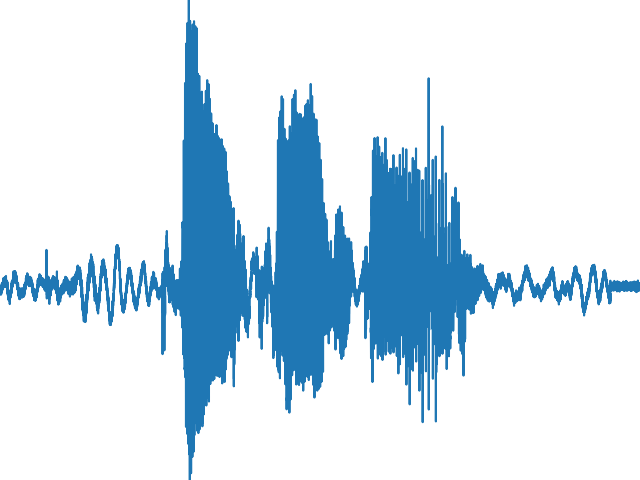}] (inputsignal) {};
\node[sum, right=\sepxx of input] (sum) {};
\node[above right=-0.1cm and \sepxxx of sum] (xt) {$\state$};
\node[myrectangle_waveform, above=\sepnodey of xt, fill overzoom image={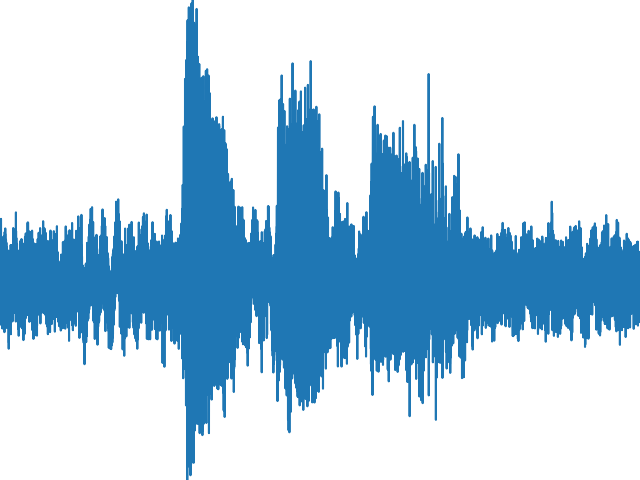}] (xtsignal) {};
\node[below=\sepy of sum] (noise) {$\boldsymbol{\epsilon}$};
\node[myrectangle3, right=\sepx of sum, align=center] (score) {Score\\Model};
\node[right=\sepxx of score] (est) {$\scomodel$};

\draw[->] (input) to (sum);
\draw[-, dashed] (inputsignal) to (input);
\draw[->] (noise) to (sum);
\draw[-, dashed] (xtsignal) to (xt);
\draw[->] (sum) to (score);
\draw[->] (score) to (est);

\end{tikzpicture}
}

\vspace{\vy}

\begin{center}
     \textbf{Latent Diffusion}
\end{center}

\scalebox{\scale}{
\begin{tikzpicture}

\coordinate (pos_x) at (1.8, 0);

\node (input) at (pos_x) {$\clean$};
\node[myrectangle_waveform, left=\sepnodex of input, fill overzoom image={images/wav_0.png}] (inputsignal) {};
\node[myrectangle3, right=\sepxx of input] (stft) {$\mathrm{Encoder}$};
\node[sumcolor, right=\sepx of stft] (sum) {};

\node[above left=-0.1cm and \sepxxx of sum] (inputstft) {$\textcolor{figdarkergreen}{\mathbf{Z}_0}$};
\node[mycircle, above left=\sepnodey and -0.5cm of inputstft, fill overzoom image={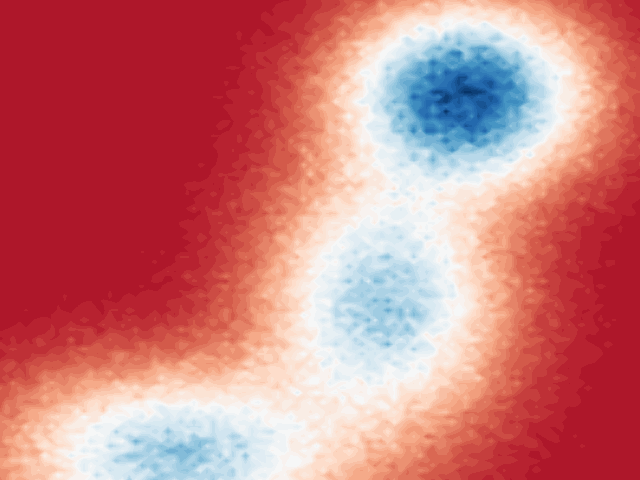}] (inputstftsignal) {};

\node[above right=-0.1cm and \sepxxx of sum] (xtstft) {$\textcolor{figdarkergreen}{\mathbf{Z}_\tau}$};
\node[mycircle, above right=\sepnodey and -0.5cm of xtstft, fill overzoom image={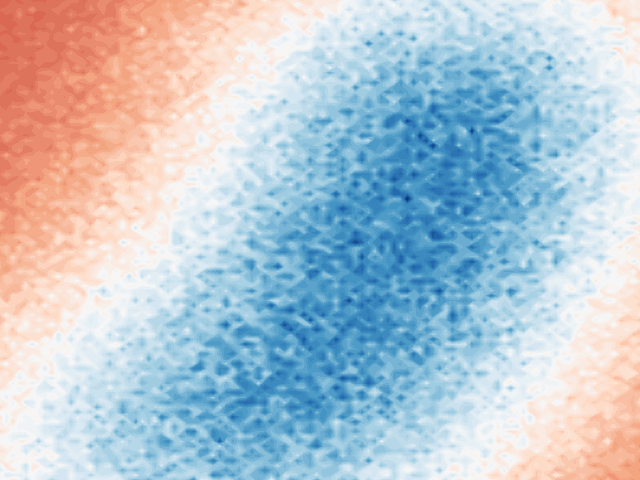}] (xtstftsignal) {};

\node[below=\sepy of sum] (noise) {$\boldsymbol{\epsilon}$};
\node[myrectangle3, draw=figdarkergreen, right=\sepx of sum, align=center] (score) {\textcolor{figdarkergreen}{Score}\\\textcolor{figdarkergreen}{Model}};
\node[right=\sepxx of score] (est) {$\textcolor{figdarkergreen}{\mathbf{s}_\theta(\mathbf{Z}_\tau, \tau)}$};

\draw[-, dashed] (inputsignal) to (input);
\draw[->] (input) to (stft);
\draw[->, figdarkergreen] (stft) to (sum);
\draw[-, dashed, figdarkergreen] (inputstft) to (inputstftsignal);
\draw[->, figdarkergreen] (noise) to (sum);
\draw[-, dashed, figdarkergreen] (xtstft) to (xtstftsignal);
\draw[->, figdarkergreen] (sum) to (score);
\draw[->, figdarkergreen] (score) to (est);

\end{tikzpicture}
}

\end{minipage}%
\hspace{0.6cm}%
\begin{minipage}{0.49\textwidth}
\begin{center}
    \textbf{STFT Diffusion }
\end{center}

\scalebox{\scale}{
\begin{tikzpicture}

\coordinate (pos_x) at (1.8, 0);

\node (input) at (pos_x) {$\clean$};
\node[myrectangle_waveform, left=\sepnodex of input, fill overzoom image={images/wav_0.png}] (inputsignal) {};
\node[myrectangle3, right=\sepxx of input] (stft) {$\mathrm{STFT}$};
\node[sumcolor, right=\sepx of stft] (sum) {};

\node[above left=-0.1cm and \sepxxx of sum] (inputstft) {$\textcolor{figdarkergreen}{\mathbf{X}_0}$};
\node[mysquare, above left=\sepnodey and -0.5cm of inputstft, fill overzoom image={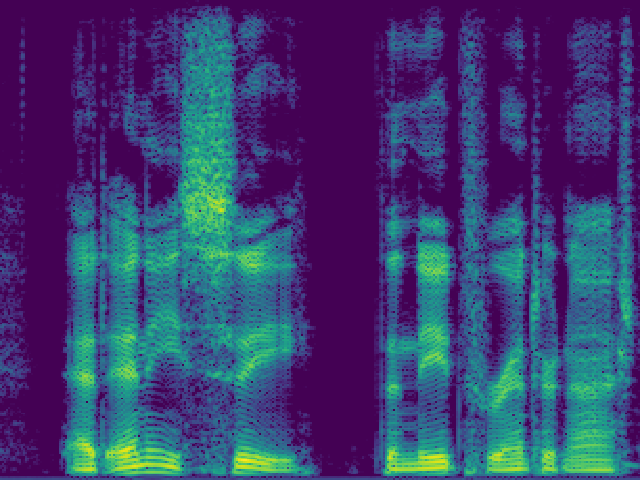}] (inputstftsignal) {};

\node[above right=-0.1cm and \sepxxx of sum] (xtstft) {$\textcolor{figdarkergreen}{\mathbf{X}_\tau}$};
\node[mysquare, above right=\sepnodey and -0.5cm of xtstft, fill overzoom image={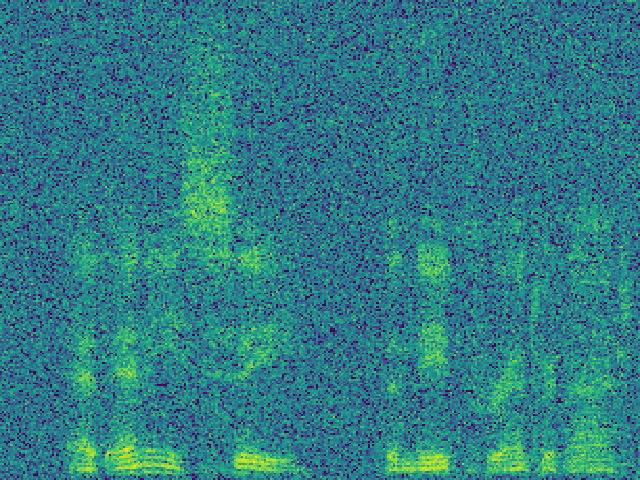}] (xtstftsignal) {};

\node[below=\sepy of sum] (noise) {$\boldsymbol{\epsilon}$};
\node[myrectangle3, draw=figdarkergreen, right=\sepx of sum, align=center] (score) {\textcolor{figdarkergreen}{Score}\\\textcolor{figdarkergreen}{Model}};
\node[right=\sepxx of score] (est) {$\textcolor{figdarkergreen}{\mathbf{s}_\theta(\mathbf{X}_\tau, \tau)}$};

\draw[-, dashed] (inputsignal) to (input);
\draw[->] (input) to (stft);
\draw[->, figdarkergreen] (stft) to (sum);
\draw[-, dashed, figdarkergreen] (inputstft) to (inputstftsignal);
\draw[->, figdarkergreen] (noise) to (sum);
\draw[-, dashed, figdarkergreen] (xtstft) to (xtstftsignal);
\draw[->, figdarkergreen] (sum) to (score);
\draw[->, figdarkergreen] (score) to (est);

\end{tikzpicture}
}

\vspace{\vy}
\begin{center}
    \textbf{CQT Score Estimation}
\end{center}
\hspace{-0.5cm}
\scalebox{\scale}{
\begin{tikzpicture}

\coordinate (pos_x) at (1.8, 0);

\node (input) at (pos_x) {$\clean$};
\node[myrectangle_waveform, left=\sepnodex of input, fill overzoom image={images/wav_0.png}] (inputsignal) {};
\node[sum, right=\sepxx of input] (sum) {};

\node[above right=-0.1cm and \sepxxx of sum] (xt) {$\state$};
\node[myrectangle_waveform, above=\sepnodey of xt, fill overzoom image={images/wav_t.png}] (xtsignal) {};

\node[below=\sepy of sum] (noise) {$\boldsymbol{\epsilon}$};

\node[myrectangle3, right=\sepx of sum] (cqt) {$\mathrm{CQT}$};

\node[above right=-0.5cm and \sepxxx of cqt] (xtcqt) {$\textcolor{figdarkergreen}{\mathbf{X}_\tau}$};
\node[mysquare, above=\sepnodey of xtcqt, fill overzoom image={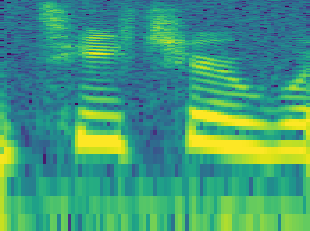}] (xtcqtsignal) {};

\node[myrectangle3, draw=figdarkergreen, right=\sepx of cqt, align=center] (score) {\textcolor{figdarkergreen}{Score}\\\textcolor{figdarkergreen}{Model}};
\node[myrectangle3, right=\sepxx of score] (icqt) {$\mathrm{iCQT}$};
\node[right=\sepxx of icqt] (est) {$\scomodel$};

\draw[-, dashed] (inputsignal) to (input);
\draw[->] (input) to (sum);
\draw[->] (noise) to (sum);
\draw[-, dashed] (xtsignal) to (xt);
\draw[->] (sum) to (cqt);
\draw[->, figdarkergreen] (cqt) to (score);
\draw[-, dashed, figdarkergreen] (xtcqtsignal) to (xtcqt);
\draw[->, figdarkergreen] (score) to (icqt);
\draw[->] (icqt) to (est);

\end{tikzpicture}
}
\vspace{-0.5cm}
\end{minipage}
    \caption{A diffusion model may be trained in (top-left) waveform \cite{kong2021diffwave}, (top-right) STFT  \cite{Welker2022SGMSE}, (bottom-left) latent  \cite{liu2023audioldm}, or (bottom-right) CQT \cite{moliner_solving_2022} domains. Sampling from the transition kernel $\kernel$ can be realized by rescaling the clean data sample $\clean$ and adding Gaussian noise $\boldsymbol{\epsilon}$ with standard deviation $\sigma(\tau)$ (see \eqref{eq:transition_kernel}). In the top-left, top-right, and bottom-left figures, the noise and the score functions are in the same domain. In the bottom-right figure, the diffusion process is formulated in the time domain but the score model pipeline includes a \ac{cqt} and its inverse.}
    \label{fig:domains}
\end{figure}

Diffusion processes can be defined for various data representations, depending on the audio application considered. Early works such as \cite{lu2022conditional, chen2021wavegrad, kong2021diffwave} directly use the waveform representation, whereas some speech enhancement approaches employ the complex \ac{stft} domain \cite{Welker2022SGMSE, richter2023speech, Lemercier2022storm}, and several music restoration works consider the \ac{cqt} domain, which is a natural space for harmonic music signals \cite{moliner2023diffusion, moliner_solving_2022, moliner2023zeroshot}. Learned domains like, e.g., autoencoder latent spaces, can also be exploited for diffusion to reduce the dimensionality of the original audio data or leverage auto-encoding properties, which gives birth to \textit{latent diffusion models} \cite{liu2023audioldm}. Figure \ref{fig:domains} offers a schematic overview of diffusion models defined in various domains. 

It should be noted that a connection between score-based diffusion models and continuous normalizing flows has been drawn by Lipman et al., creating a new category of so-called \emph{flow matching} models \cite{lipman2023flow}. Flow matching methods generalize Gaussian diffusion models and allow to design more flexible probability paths (based on, e.g., optimal transport) between arbitrary terminal distributions. This approach has been used to train a foundational speech model, which can be finetuned to perform restoration tasks such as speech enhancement and separation \cite{liu2024generative}.

\subsection*{Model-based processing with diffusion models}
\label{sec:modelbased}

In statistical model-based speech enhancement, each time-frequency bin of the speech and noise spectrograms is often assumed to be mutually independent and to follow a zero-mean complex Gaussian prior distribution \cite{gerkmann2018book_chapter}. For an additive mixture model, this yields a Gaussian likelihood model for the mixture and a Gaussian posterior model for the clean speech estimate using Bayes' rule. Under this Gaussian assumption, the posterior mean can be derived as the celebrated Wiener filter solution, providing the optimal speech estimate in the \ac{mmse} sense. However, distributional and independence assumptions are merely approximations utilized out of convenience for the derivation of closed-form estimators, e.g., the mentioned Wiener filter. With diffusion models, there are no distributional and independence assumptions on the speech and noise signals themselves. Indeed, the very intent of deep generative modeling is to allow more flexibility by inferring the signal structure from data rather than the parameters of a fixed distribution.

In contrast to other deep learning approaches to audio restoration, two aspects make diffusion models well suited for the introduction of domain knowledge, showcasing them as model-based approaches. The first property is derived from the physical inspiration of diffusion models and their connection to Gaussian denoising~\cite{song2019generative}, which makes them easier to interpret in comparison to other deep generative models such as \acp{gan}. In particular, the Gaussian parameterization of the transition kernel $\kernel$ enables the injection of knowledge in the form of specific schedules for the mean $\boldsymbol{\mu}$ and standard deviation $\sigma$ \cite{Welker2022SGMSE, lu2022conditional, lay2023reducing}. Furthermore, domain knowledge can be leveraged to posit a distributional hypothesis for the noise process $\wien$ used during forward and reverse diffusion. For instance, Nachmani et al.~\cite{nachmani2022gamma} propose a Gamma distribution instead of the usual Wiener process $\wien$ with Gaussian increments, as it better fits the estimation error distribution. The authors consequently show improvements in speech generation quality compared to the Gaussian case.

The second powerful property of diffusion models is their natural integration within stochastic optimization and posterior sampling using Bayes' theorem, making them particularly suited for conditional generation. We consider the case of audio restoration under the scope of inverse problem solving, i.e., retrieval of clean audio $\clean$ from a measurement $\y$. There, an approximation of the measurement likelihood $p(\y | \clean)$ can be obtained via a closed-form model of the operation corrupting $\clean$ into $\y$. Combining this likelihood model and the learned deep generative prior with Bayes' rule can provide sampling or stochastic optimization algorithms for the conditional generation of samples from the posterior distribution $p(\clean | \y)$ \cite{chung_diffusion_2022, lemercier2023derevdps, moliner_solving_2022, moliner2023zeroshot}.

In summary, first, we see that the data-driven nature of diffusion models allows a higher degree of versatility than traditional signal processing methods, which are often strictly based on simple closed-form distributions and independence assumptions. Secondly, it is important to note that diffusion models transcend the stereotype of being non-interpretable black-boxes. Instead, they benefit from strong integration within stochastics and enable significant potential for the injection of domain knowledge for model-based audio processing.

\section*{Conditional Generation with Diffusion Models}
\label{sec:cond}
One of the most fundamental uses of diffusion models is to perform unsupervised learning from a finite collection of samples to learn an underlying complex data distribution. This provides the ability of \emph{unconditional generation}, i.e., to generate new samples from the learned data distribution. To solve audio restoration tasks, a diffusion model must be adapted to generate audio that not only conforms to the learned clean audio distribution but, importantly, is also a plausible reconstruction of a given corrupted signal. This effectively requires the diffusion model to perform \emph{conditional generation}. We distinguish between three families of approaches for diffusion-based generative audio restoration: \emph{(i) input conditioning}, where the score model is provided with a task-specific conditioning signal as input, \emph{(ii) task-adapted diffusion}, where the forward and reverse diffusion processes are modified to interpolate between clean and corrupted signals, and \emph{(iii) external conditioning}, where the score model is trained purely on clean audio data and is later combined with an external conditioner during inference. Approaches that use input conditioning (i) or external conditioning (iii) often initialize the iterative generation process with pure Gaussian noise, and then generate a clean signal by iteratively filtering this noise while being guided by the conditioning signal. In contrast, in task-adapted diffusion (ii), the corrupted audio itself is used for initialization and iteratively filtered, making this approach conceptually closer to a denoising procedure.

\subsubsection*{(i) Input conditioning}

Diffusion models that use input conditioning are provided with a task-specific conditioning signal~$\cond$ (usually some representation of the corrupted signal $\y$) as an additional input during training and inference. To this end, they employ \acp{dnn} as score models that are specifically designed to perform feature fusion between the inputs $\state$ and $\cond$. It should be noted that, in most cases, input conditioning approaches require the use of paired data, as the conditioning signal~$\cond$ and the target data sample $\mathbf x_0$ should be representations of the same data instance, or at least share some semantics. The earliest works to follow this approach include DiffWave \cite{kong2021diffwave}, which uses mel-spectrograms as conditioning signals for neural vocoding and text-to-speech tasks. While DiffWave focuses on audio generation rather than restoration, the authors of DiffWave also provide preliminary evidence that an unconditional speech diffusion model can perform speech enhancement by using the corrupted audio $\y$ as a starting point of the sampling process even though the diffusion model was only trained to remove Gaussian noise. DiffuSE \cite{lu2021study} builds upon DiffWave to solve speech enhancement tasks, using noisy spectral features as conditioning $\cond$.

In the worst case, the score model may not use conditioning $\cond$ at all, thus inadvertently performing unconditional rather than conditional generation. One possible solution to this is \emph{classifier-free guidance}, where the conditioning signal is randomly set to zero with a fixed probability during training. This results in a single model that can  both provide an estimate for the conditional score $\nabla_{\state} \log p_\tau(\state | \cond)$ and the unconditional score $\sco$. At inference, the two estimates can then be weighted at will to trade quality (more weight on conditional score) for variety (more weight on unconditional score). This idea has been used, for instance, by Liu et al.~\cite{liu2023audioldm} to perform controllable full-band audio synthesis and can also be employed for various audio restoration tasks.

\subsubsection*{(ii) Task-adapted diffusion}

\begin{figure}
     \scalebox{0.87}{
     \input{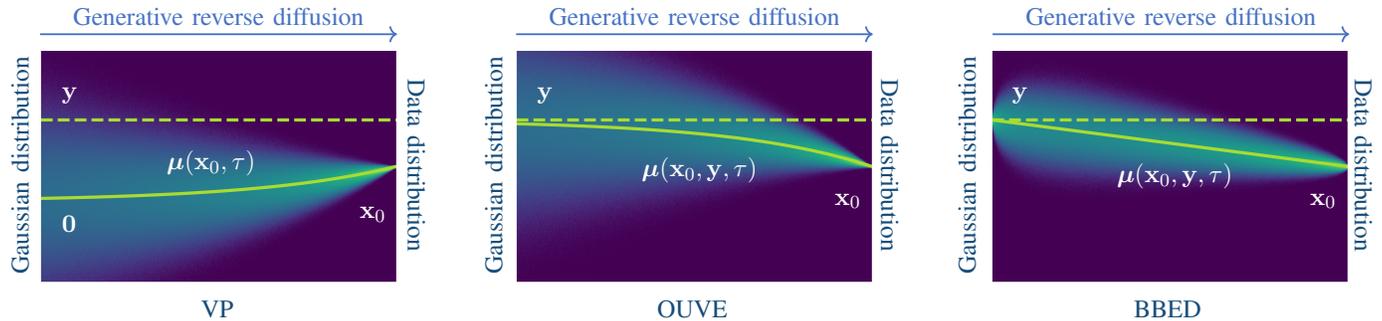}
     }
     \caption{Comparison of different diffusion processes. (left) Classical variance-preserving (VP) diffusion: mean exponentially interpolates between clean audio $\clean$ and $\mathbf{0}$, irrespectively of the degraded audio $\y$ \cite{song2021sde}. (middle) Task-adapted Ornstein–Uhlenbeck variance exploding (OUVE) diffusion: mean exponentially interpolates between clean audio $\clean$ and degraded audio $\y$\cite{Welker2022SGMSE, richter2023speech}. (right) Task-adapted brownian bridge with exponential diffusion (BBED): mean linearly interpolates between clean audio $\clean$ and degraded audio $\y$\cite{lay2023reducing}.}
        \label{fig:trajectories}
\end{figure}

In many restoration tasks such as denoising, dereverberation and separation, the corrupted signal $\y$ and the clean signal $\clean$ have same dimensionality. This allows one to define what we denote as \textit{task-adapted} diffusion processes, i.e., diffusion processes whose mean $\boldsymbol{\mu}(\clean, \y, \tau)$ is $\clean$ at $\tau=0$ and $\y$ at $\tau=T$, and that interpolates between these terminal values for $\tau \in [0, T]$.  This is a form of conditioning which is not introduced as an auxiliary variable to the score model $\mathbf{s}_\theta$ as for input conditioning, but rather directly injected in the parameters of the diffusion process itself. Examples of classical and task-adapted diffusion processes are visualized on Figure~\ref{fig:trajectories}. CDiffuSE \cite{lu2022conditional} is one of the earliest methods using task-adapted diffusion, formulating the process in discretized time steps. \ac{sgmse} \cite{Welker2022SGMSE} and \ac{sgmse}+ \cite{richter2023speech} extend this idea to the continuous \ac{sde}-based formalism of diffusion models to derive pairs of forward and backward processes. Subsequent works \cite{Lemercier2022storm,lay2023reducing} build upon this formalism to design alternative forward and backward processes which result in fewer sampling steps and/or higher reconstruction quality. In practice, these methods combine task-adapted diffusion processes with input conditioning, by also providing $\y$ as an auxiliary input to the score model.  

The interpolation between $\clean$ and $\y$ underlying these approaches assumes an additive signal model typical for denoising tasks, which can also be treated as natural for separation tasks \cite{scheibler2022diffusionbased} or for convolutive corruptions such as reverberation. The aforementioned methods also achieve excellent reconstruction quality for non-additive corruptions like in bandwidth extension \cite{Lemercier2022storm} and STFT phase retrieval \cite{peer2023diffphase}, which shows their ability to perform \emph{blind} restoration, i.e., when the corruption operator is not perfectly known during inference.

\subsubsection*{(iii) External conditioning}

External conditioning approaches combine an unconditional diffusion model with an external conditioner that provides a conditioning signal during inference. Since the diffusion model is unconditional, no knowledge of the restoration task is accessed during the training stage and no supervision nor paired data is required. Instead, the task-specific information is injected only at inference by the external conditioner. Therefore, external condition methods can leverage diffusion-based foundation models pre-trained on large-scale data, and adapt them for inference without further re-training. One such type of external conditioner is a pre-trained classifier enabling the combined model to perform class-conditional data generation. For audio restoration, the external conditioner usually takes the form of a task-specific closed-form measurement model. This results in an overall model that combines a strong data-driven prior for clean audio (score model) with a model-based formulation of the specific restoration task (measurement model). This approach shows good results even when the observation $\y$ is affected by measurement noise \cite{moliner_solving_2022,lemercier2023derevdps} and has the advantage of not requiring retraining of the diffusion model for new restoration tasks. These approaches can be applied to blind restoration tasks if a good parameterization of the measurement operator is found. The parameterization enables classical estimation algorithms to be utilized for joint inference of the measurement model and target audio sample estimation, as Moliner et al.~\cite{moliner2023zeroshot} accomplished in the blind bandwidth extension of historical music recordings.

\section*{Diffusion Models for Inverse Problems}
\label{sec:inverse-problems}
\begin{figure}
    \centering
    \input{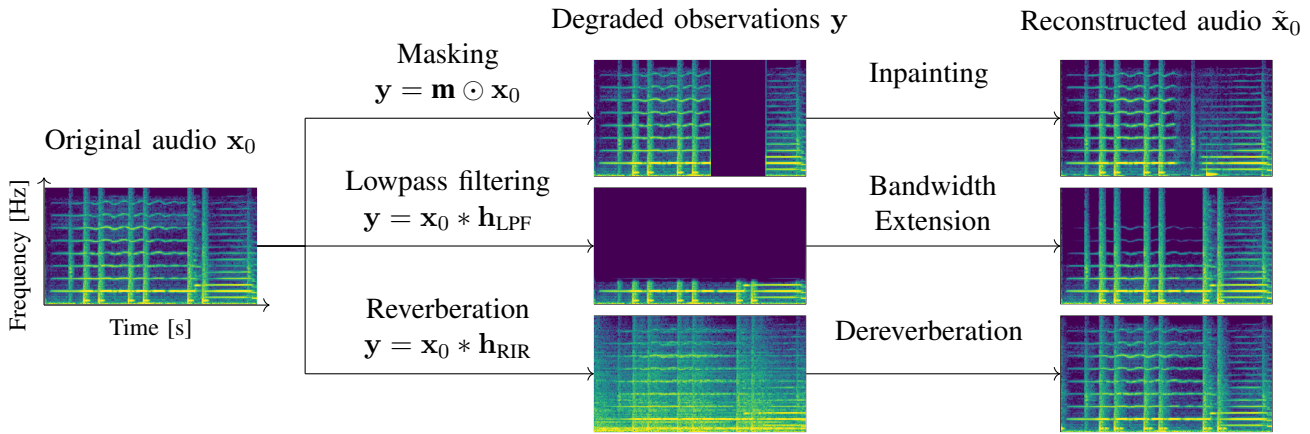}
\caption{ Visual representation of several inverse problems in audio: (top to bottom) inpainting, bandwidth extension, and dereverberation. (Left) The spectrogram of the original audio signal, \(\clean \), undergoes various transformations via different measurement operators, (middle) the resulting degraded observations, \(\y\), correspond to specific audio distortions, and (right) the reconstructed audio signal \(\tilde{\x}_0\) is obtained by solving each inverse problem. Notably, the reconstructed example spectrograms (right) closely mirror the original (left), but minor differences appear because of the inherent ill-posed nature of these inverse problems.}
    \label{fig:inverse problems}
\end{figure}

We have seen different strategies to condition diffusion models for audio restoration tasks. This section delves into the \emph{external conditioning} approach, specifically focusing on the application of diffusion models for solving inverse problems in the audio domain. Several audio restoration tasks can be formulated as an inverse problem, wherein an observed audio signal $\y$ is the result of corrupting a clean signal  $\clean$ with a degradation model $\mathcal{A}(\cdot)$ and additive noise $\n$, which can be expressed as 
\begin{equation} \label{eq:measurement}
\y=\mathcal{A}(\clean)+\n.
\end{equation}
This model covers an infinite set of possible degradations, depending on how the operator $\mathcal{A}(\cdot)$ is defined. Three cases of particular interest are showcased in Figure \ref{fig:inverse problems}. Initially, we concentrate on scenarios in which both the degradation model $\mathcal{A}(\cdot)$ and the noise statistics $\n$ are known. The goal is to recover the original signal $\clean$ from the corrupted observations $\y$. However, in many cases, the problem is ill-posed, lacking a unique solution and defying straightforward resolution.

Often, solving an inverse problem is approached with a \ac{map} objective 
\begin{equation}
\label{eq:inverse_problem_MAP}
    \underset{\clean}{\mathrm{arg\,max}}\;  
    p(\clean | \y)\,,
\end{equation}
where the posterior distribution factorizes into likelihood and prior $p(\clean | \y) \propto p(\y|\clean) p(\clean) $. Under a zero-mean Gaussian measurement noise assumption, denoted as $\mathbf n \sim \mathcal{N}(\mathbf{0}, \sigma^2_y \mathbf{I})$, the \ac{map} estimate takes the form
\begin{equation}
\label{eq:inverse_problem_optim}
    \underset{\clean}{\mathrm{arg\,min}}\;  
    \displaystyle{\frac{1}{\sigma_y^2}}
     \lVert \y -  \mathcal{A}(\clean)
    \rVert_2^2
    + \mathcal{R}(\clean)\,,
\end{equation}
where the first term is a reconstruction cost function, in this case an $L^2$-norm, designed to preserve fidelity with the observations $\y$. The second term, $\mathcal{R}(\clean)$, functions as a regularizer, incorporating prior information or domain knowledge about the signal. Its purpose is to mitigate the under-determination of the problem by constraining the space of suitable solutions, thereby making the optimization feasible in practice. In audio processing, a frequently employed regularizer is the sparsity-promoting $L^1$-norm, which assumes that the true signal is sparse in a specified transform domain, such as time-frequency representations.

A diffusion model learns the statistical characteristics of the training data, in our case of clean audio signals. One can then expect diffusion models to have the potential to serve as strong data-driven priors for solving inverse problems. We will now elaborate on how to leverage these diffusion-based generative priors for solving \eqref{eq:inverse_problem_optim}.

To solve an inverse problem using a diffusion model, the score $\sco$ in the reverse \ac{sde} \eqref{eq:reverse-sde} is replaced with the score of the posterior using Bayes' rule
\begin{equation}
     \postsco = \sco +  \likelihood \,,
\end{equation}
where the \emph{prior score} $\sco$ is approximated with the unconditional score model $\mathbf{s}_\theta$ (see \eqref{eq:training}). The term $\likelihood$ represents the \emph{likelihood score}. However, it is important to note that the likelihood $p(\y | \state)$ is only analytically tractable for $\tau=0$, as $\state$ refers to a noisy version of $\clean$ and the true likelihood is defined through an intractable integral over all possible $\clean$
\begin{equation}
   p_\tau(\y | \state)= \int_{\clean} p(\y | \clean) p_\tau(\clean | \state) \diff{\clean} \,.
\end{equation}

Some works alleviate this issue by simply bypassing the likelihood term, and instead project the state $\state$ onto the set of the observations $\y$ at every step of the discretized inference process \cite{song2021sde}. The objective of such \textit{projection-based} method is to inject the reliable parts of the observations into the intermediate predictions. This ensures that at each step, the intermediate output of the algorithm is consistent with the algorithm input, i.e., the degraded audio, which is often refered to as \textit{data consistency} and helps avoiding degenerate solutions. Projection-based methods offer the advantage of ensuring data consistency and simplicity in terms of algorithmic implementation. However, their applicability is limited to a reduced set of linear inverse problems, such as audio inpainting or bandwidth extension \cite{liu2023audioldm, moliner_solving_2022}, where closed-form expressions for the projection step are available.

\begin{figure}
    \centering
    \input{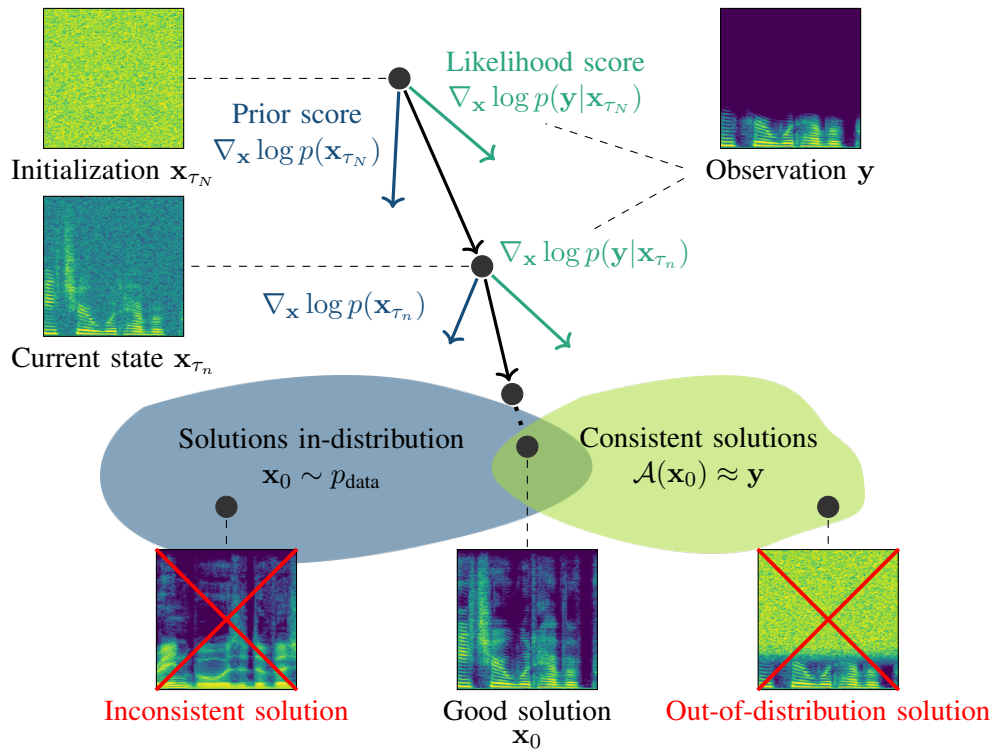}
    \caption{Geometrical interpretation of posterior sampling with diffusion models (e.g., \cite{chung_diffusion_2022}). The prior score guides the trajectories towards solutions within the training data manifold, or in-distribution with the training data (gray space). Simultaneously, the role of the likelihood score is to steer the sampling trajectories toward a solution space consistent with the observed data (light green space). When properly weighted, the two components pull the sampling process to the intersection of these two manifolds. This intersection exists and contains the solutions to the inverse problem if the two score functions are properly estimated and if the solutions are contained in the manifold spanned by the training data, i.e., if the training dataset is properly adapted to the problem.
}
    \label{fig:posterior_sampling}
\end{figure}

Other works adopt more theoretically grounded approximations of the likelihood that allow a broader versatility by incorporating a model-based approach.  In particular, Chung et al. \cite{chung_diffusion_2022} proposed Diffusion Posterior Sampling, and approximate the likelihood as $p_\tau(\y | \state) \approx p(\y | \hat{\mathbf{x}}_0 (\state) )$. There, $\hat{\mathbf{x}}_0 (\state)$ is a coarse estimate of $\clean$ obtained by denoising from state $\state$ in just one deterministic reverse diffusion step. When modeling the measurement noise $\n$ in \eqref{eq:measurement} as a Gaussian $\n \sim \mathcal{N}(\mathbf{0}, \sigma_y^2 \mathbf{I} )$, the resulting approximated likelihood is a Gaussian distribution $p \left(\y | \hat{\mathbf{x}}_0 (\state) \right) = \mathcal{N}(\mathcal{A}(\hat{\mathbf{x}}_0 (\state) ), \sigma_y^2 \mathbf{I})$. It follows that the likelihood score can be computed as:
\begin{equation} \label{eq:likelihood}
    \likelihood \approx - \frac{1}{\sigma_y^2} \nabla_{\state} 
     \lVert 
     \y - \mathcal{A} \left(
    \hat{\x}_0(\state) \right)
 \rVert^2_2\,.
\end{equation}
The $L^2$-norm in \eqref{eq:likelihood} can be replaced by any other objective function that better fits the statistics of the measurements \cite{chung_diffusion_2022}. For example in \cite{moliner2024buddy}, the measurement noise $\n$ is modelled as a Gaussian in the compressed STFT domain, where the compression is a square-root power law on the STFT magnitude, leaving the phase unchanged. This helps accounting for the heavy-tailedness of speech distributions \cite{gerkmann2010}. It is important to note that the gradient operator $\nabla_{\state}$ requires differentiating through the score model by backpropagation, which introduces a computational overhead. In practice, the unknown measurement noise variance $\sigma_y^2$ is estimated empirically using, e.g., the norm of the gradients in \eqref{eq:likelihood} \cite{moliner_solving_2022}. Compared to projection-based methods, this approach is not limited to linear problems and can be applied to cases where $\mathcal{A}(\cdot)$ is nonlinear, as long as the operator $\mathcal{A}(\cdot)$ is differentiable. A geometrical perspective on the sampling process is displayed in Figure~\ref{fig:posterior_sampling}. This diagram illustrates the intuition behind conditional sampling with a diffusion model, in this case in the context of bandwidth extension. This strategy has been successfully applied in audio bandwidth extension \cite{moliner_solving_2022}, audio inpainting \cite{moliner2023diffusion}, and dereverberation \cite{lemercier2023derevdps}.

\subsection*{Blind inverse problems}

Until this point, our analysis has proceeded under the assumption that the degradation operator $\mathcal{A}(\cdot)$ is known. However, in practical applications, the degradation operator is often unknown. This lack of knowledge about the degradation operator renders the calculation of the posterior $p(\clean | \y)$ a \emph{blind} inverse problem, substantially raising the difficulty of the task. The Diffusion Posterior Sampling approach \cite{chung_diffusion_2022}, as previously explained, provides a valuable foundation that can be extended to tackle blind inverse problems. In scenarios where we possess at least some knowledge of the structure of the degradation operator, a viable strategy is to embrace a model-based approach. This involves designing a parametric model of the degradation operator, denoted as $\mathcal{A}_\phi(\cdot)$, and jointly optimizing its parameters $\phi$ alongside the restored audio signal throughout the sampling process.

\begin{figure}
    \hspace{-1.7cm}
    \scalebox{0.985}{
    \input{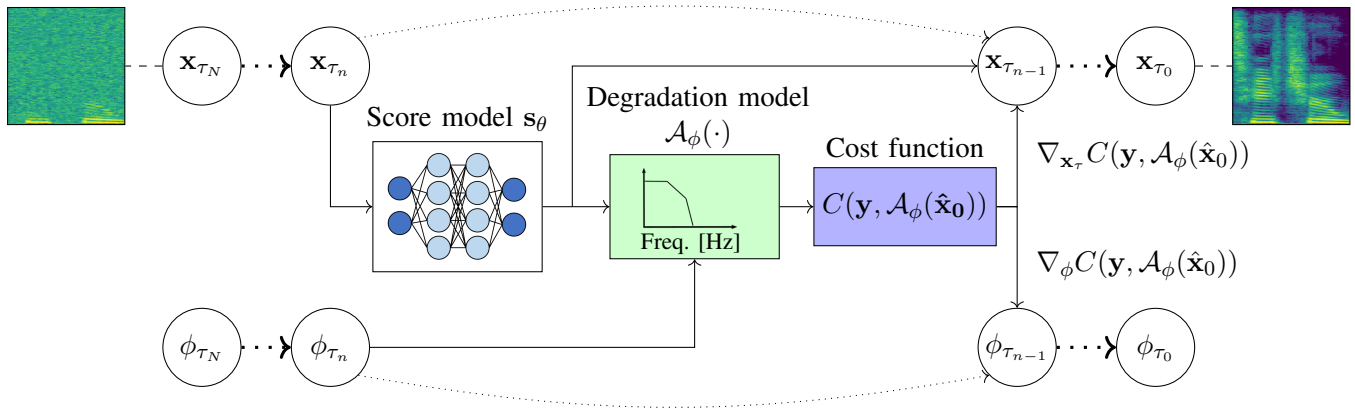}
    }
    \caption{BABE: posterior sampling algorithm for solving blind bandwidth extension using a prior score model $\mathbf{s}_\theta$ and a parameterized degradation operator $\mathcal{A}_\phi$ \cite{moliner2023zeroshot}. The optimization alternates between updating the reconstructed signal $\x$ (top) and the degradation parameters $\phi$ (bottom). For ease of reading, we write $\hat{\mathbf{x}}_0 := \hat{\mathbf{x}}_0(\mathbf{x}_{\tau_n})$.}
    \label{fig:babe}
\end{figure}

An example of this approach is the Blind Audio Bandwidth Extension (BABE) \cite{moliner2023zeroshot}, which addresses the problem of blind reconstruction of missing high frequencies in music from bandlimited observations without knowledge of the lowpass degradation, such as the cutoff frequency. This challenge is typical in restoring historical audio recordings. In BABE, the measurement model $\mathcal{A}_\phi(\cdot)$ is parameterized by a piecewise approximation of a low-pass filter in the frequency domain, where the parameters $\phi$ represent the cutoff frequencies and decay slopes of this filter \cite{moliner2023zeroshot}. The optimization process,  as illustrated in Figure \ref{fig:babe}, alternates between sampling updates of the audio signal $\state$ and refining $\phi$ through stochastic gradient descent, using a maximum likelihood objective as the guiding principle. BUDDy \cite{moliner2024buddy} takes a similar approach and solves joint speech dereverberation and room acoustics estimation by combining Diffusion Posterior Sampling with a model-based subband filter approximating room impulse response. The resulting method largely outperforms other blind unsupervised dereverberation methods. Thanks to unsupervised learning, BUDDy seamlessly adapts to new acoustic scenarios, whereas supervised methods typically struggle when there is a mismatch between training and testing conditions.

\section*{Practical Requirements of Diffusion-based Sampling for Audio Tasks}
\label{sec:practical}
While diffusion models provide powerful priors that can be employed for various audio restoration tasks, they require some improvements to be suitable for real-time acoustic communications. We divide these requirements into two categories: \textit{(i) inference speed and causal processing}, which can be prohibitive for low-latency real-time applications, and \textit{(ii) robustness to adverse conditions}, which must be assured for integration into reliable systems.

\subsection*{Inference speed and causal processing}

One major drawback of diffusion models is their slow inference. As the score model is called at each step of the reverse process, the computational complexity is directly proportional to the number of steps used and the order of the solver, i.e., the number of score estimations used per time step. Using more diffusion steps naturally provides better sample reconstruction since the truncation error of the numerical solver is reduced when the step size is decreased. Similarly, increasing the solver order reduces the per-step truncation error. However, both these options lead to an increased computational cost. Furthermore, accumulating truncation errors over the diffusion trajectory can make the samples diverge from the distribution learned during training, and therefore make the score model produce unreliable estimates, which is referred to as the \textit{drifting bias}. These two sources of error compound over the diffusion trajectory, therefore, without further optimization, high-quality reconstruction can only be obtained at a high computational cost. This section presents several methods to reduce the computational complexity of diffusion-based methods in audio applications.

\textit{Reducing per-step inference time}: A natural way to accelerate inference is to reduce the cost of each call to the score model. This can be obtained by minimizing the size of the neural network used for score inference through, e.g., knowledge distillation, or by reducing the size of the space itself where diffusion is performed, resulting in \textit{latent diffusion models}. The latent space should be designed such that its reduced dimensionality has a limited impact on the reconstruction quality, and its structure allows for score estimation with a reasonably-sized neural network. Latent diffusion is popular in text-to-audio generation and has been recently applied to audio editing (including restoration) in AUDIT \cite{Wang2023AUDITAE}, which uses latents provided by a variational autoencoder.

\begin{figure}[H]
\hspace{-0.7cm}
\scalebox{0.98}{
\newcommand{\spectrow}{.32\textwidth}
\newcommand{\spectroh}{.27\textwidth}
\newcommand{\spectroxs}{.015\textwidth}
\newcommand{\xmax}{6}

\begin{tikzpicture}[scale=0.95, transform shape]
 \protect\centering

\begin{axis}
[
    name={clean},
    title = {Anechoic},
    axis line style={draw=none},
    xmin = 0, xmax = \xmax,
    ymin = 0, ymax = 8000,
    xtick = {0, 1, 2, 3, 4, 5},
    xticklabels=\empty,
    yticklabels=\empty,
    xlabel = {Time [s]},
    ylabel = {Frequency [Hz]},
    width =\spectrow,
    height =\spectroh,
    title style={yshift=-5pt},
    xlabel style={yshift=5pt},
    ylabel style={yshift=-5pt},
]
\addplot graphics[xmin=0,ymin=0,xmax=\xmax,ymax=8000,
includegraphics={trim=2cm 0cm 5cm 0cm,clip}] 
{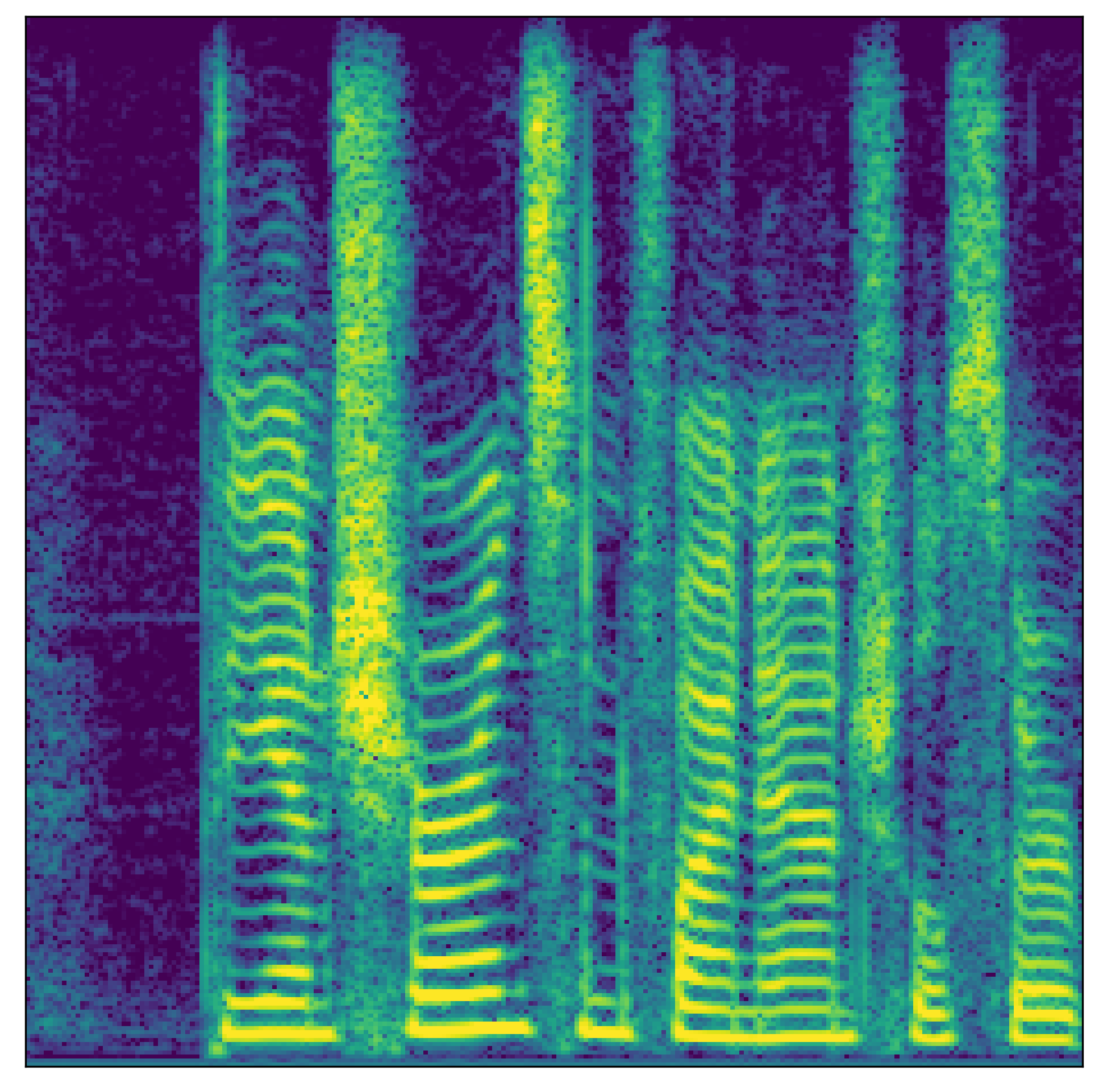};
\end{axis}

\begin{axis}
[
    name = {noisy},
    title = {Reverberant},
    at={(clean.south east)},
    xshift = \spectroxs,
    axis line style={draw=none},
    xmin = 0, xmax = \xmax,
    ymin = 0, ymax = 8000,
    xticklabels=\empty,
    yticklabels=\empty,
    xlabel = {Time [s]},
    width =\spectrow,
    height =\spectroh,
    title style={yshift=-5pt},
    xlabel style={yshift=5pt},
]
\addplot graphics[xmin=0,ymin=0,xmax=\xmax,ymax=8000,
includegraphics={trim=2cm 0cm 5cm 0cm,clip}] 
{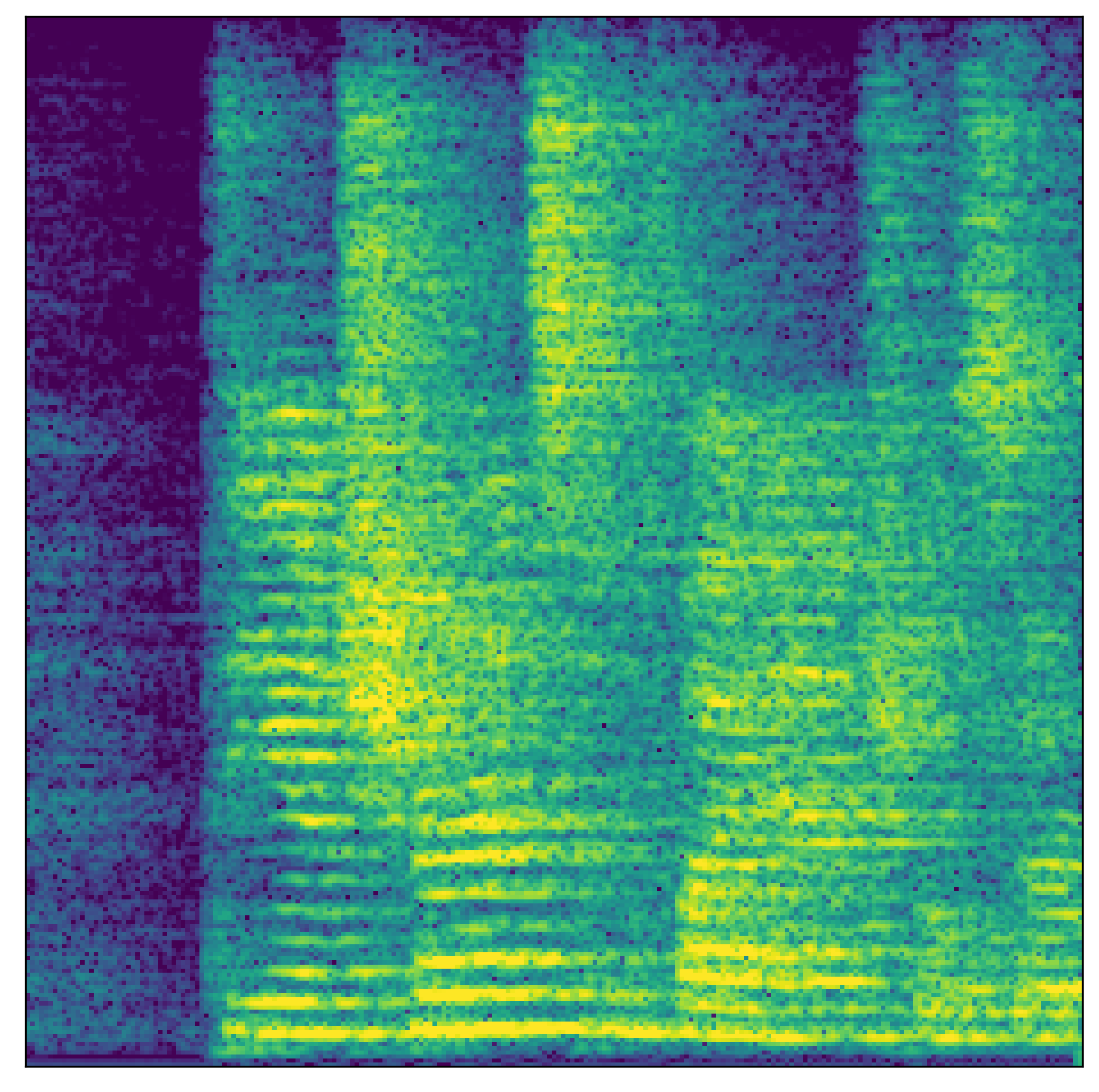};
\end{axis}

\begin{axis}
[
    name = {ncsn},
    title = {Initial Prediction},
    at={(noisy.south east)},
    xshift = \spectroxs,
    axis line style={draw=none},
    xmin = 0, xmax = \xmax,
    ymin = 0, ymax = 8000,
    xticklabels=\empty,
    yticklabels=\empty,
    xlabel = {Time [s]},
    width =\spectrow,
    height =\spectroh,
    title style={yshift=-5pt},
    xlabel style={yshift=5pt},
]
\addplot graphics[xmin=0,ymin=0,xmax=\xmax,ymax=8000,
includegraphics={trim=2cm 0cm 5cm 0cm,clip}] 
{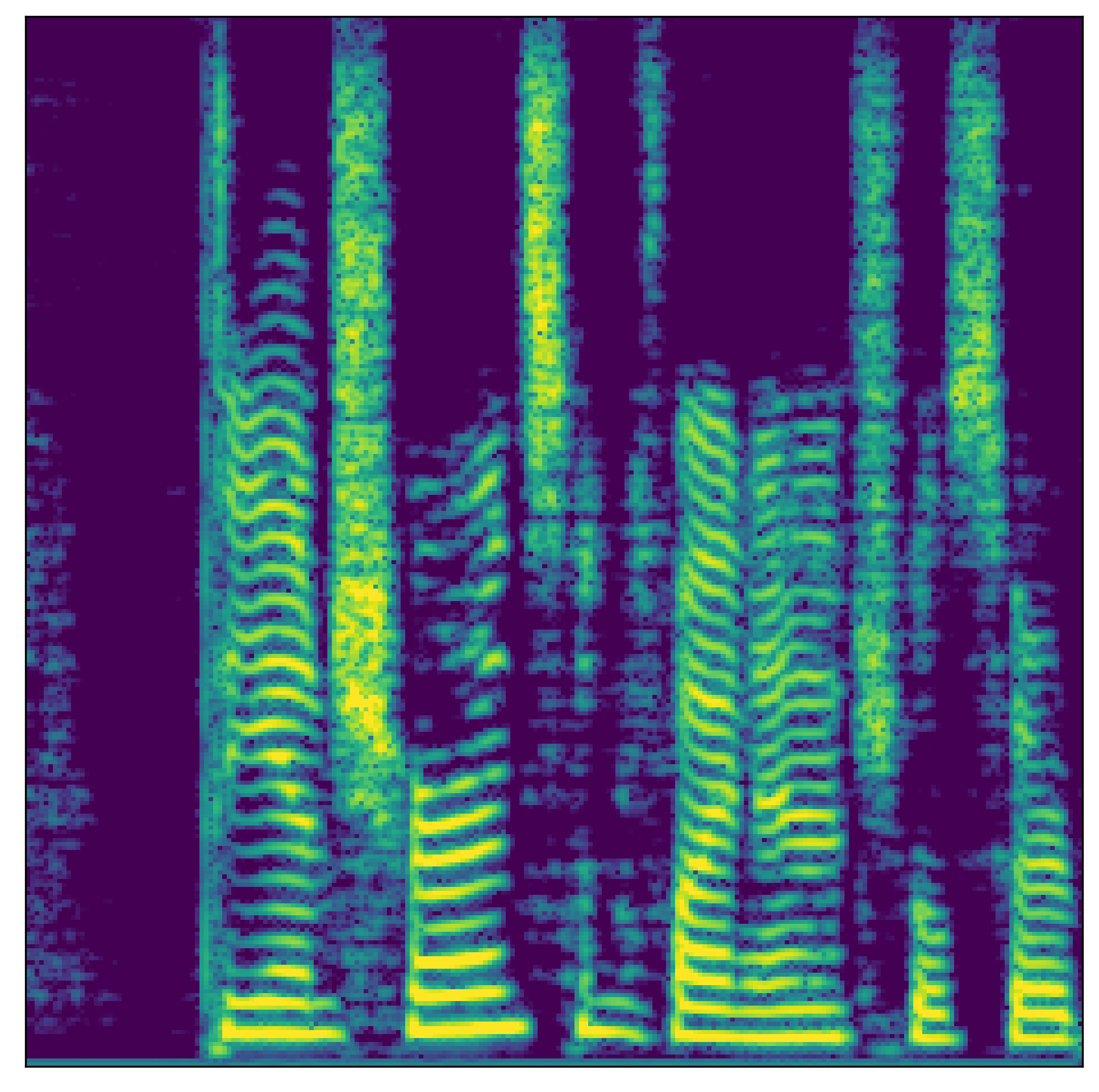};

\end{axis}

\begin{axis}
[
    name = {regensgmse},
    title = {StoRM Restoration},
    at={(ncsn.south east)},
    xshift = \spectroxs,
    axis line style={draw=none},
    xmin = 0, xmax = \xmax,
    ymin = 0, ymax = 8000,
    xticklabels=\empty,
    yticklabels=\empty,
    xlabel = {Time [s]},
    width =\spectrow,
    height =\spectroh,
    title style={yshift=-5pt},
    xlabel style={yshift=5pt},
]
\addplot graphics[xmin=0,ymin=0,xmax=\xmax,ymax=8000,
includegraphics={trim=2cm 0cm 5cm 0cm,clip}] 
{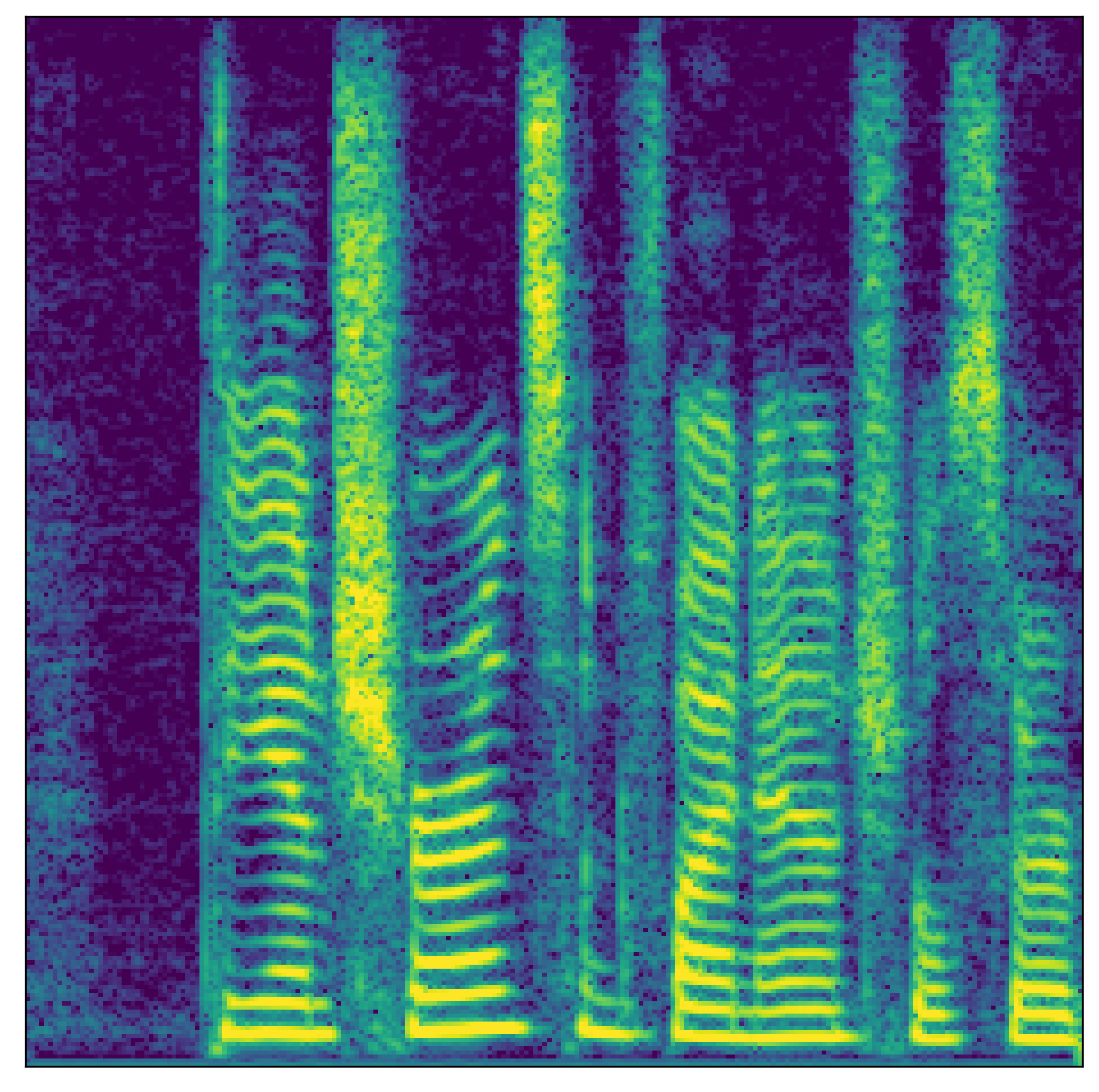};

\end{axis}

 \begin{axis}
[
    at={(regensgmse.south east)},
    xshift = 0.01\textwidth,
    yshift = -0.021\textwidth,
    width = 0.13\textwidth,
    height = \spectroh + 0.78cm,
    hide axis,
]
\addplot graphics[xmin=0,ymin=0,xmax=1,ymax=1] {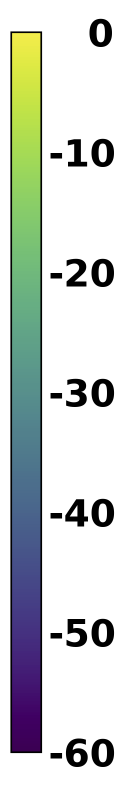};
\end{axis}

\end{tikzpicture}
 
}
\caption{Dereverberation results with StoRM \cite{Lemercier2022storm}. Input $T_\mathrm{60}$ is 1.06~$\mathrm{s}$. Three seconds of audio are shown, and the bandwidth is 8 kHz. Severe speech distortions are observed in the initial prediction because of the harsh reverberant conditions. StoRM corrects the distortions and restores the formant structure without residual reverberation.}
 \vspace{-1em}
\label{fig:spectros}
\end{figure}

\textit{Improving initialization}: Another possibility to accelerate sampling is to find a better initial prediction to reduce the distance between the initial condition $\init$ and the target sample $\clean$. This can be provided by a separate plug-in predictive network providing an estimate of the posterior mean $\mathbb{E}[\clean | \y]$ as proposed by Lemercier et al. in their Stochastic Regeneration Model (StoRM)~\cite{Lemercier2022storm} for speech enhancement (see Figure~\ref{fig:storm}). The diffusion-based generative model can restore target cues potentially destroyed by the predictive stage while additionally removing residual corruption. The resulting approach requires significantly fewer function evaluations than the original diffusion-only model in \cite{richter2023speech}, for a better-sounding result.  Figure~\ref{fig:spectros} shows the clean, degraded, and restored speech spectrograms produced with StoRM. As a simpler alternative, the corrupted utterance $\y$ can be directly used as the mean of the initial state $\init$. This latter strategy is sometimes referred to as \textit{warm initialization} and has already been used in audio-related tasks such as speech enhancement \cite{lu2022conditional} and bandwidth extension \cite{moliner2023zeroshot}. A good initial prediction can also be obtained by designing a more suitable diffusion trajectory to reduce the mismatch between training and inference, as suggested by Lay et al.~\cite{lay2023reducing} for speech enhancement. As shown in Figure~\ref{fig:trajectories}, the BBED diffusion process proposed in \cite{lay2023reducing} has a linear, constant speed mean interpolating between the clean and noisy speech, which effectively terminates at the clean speech in finite time, unlike the original OUVE diffusion process proposed in \cite{Welker2022SGMSE}. 

\begin{figure}[t]
    \input{figures/clouds.tex}
    \caption{Visualization of the inference process for the predictive, generative and StoRM \cite{Lemercier2022storm} models for a complex posterior distribution. 
    With the proposed two-stage inference, StoRM uses the predictive mapping to the posterior mean $\mathbb{E}[\clean | \y]$ as an intermediate step for 
    generation of a sample which is more likely to lie in high-density regions of the posterior $p(\clean | \y)$.}
    \label{fig:storm}
\end{figure}

\textit{Reducing the number of steps}: The remaining approaches investigate how to reduce the number of diffusion steps of the reverse process. As in most ODE/SDE integration problems, using off-the-shelf higher-order samplers can improve the per-step precision but here it comes at the cost of more calls to the neural network for each step, which leads to a non-trivial tradeoff between computational complexity and sample quality. In denoising diffusion implicit models (DDIM) \cite{song2022denoising} instead, the Markovian property of the transition kernel is deliberately removed by conditioning the next reverse diffusion estimate $\x_{\tau_{n-1}}$ on both the previous state $\x_{\tau_n}$ and $\hat{\mathbf{x}}_0$, a coarse estimate of the clean signal obtained via one-step denoising (see the section above on inverse problems). This allows an arbitrary number of steps to be skipped during reverse diffusion, which can significantly accelerate inference. 

A progressive distillation method for reverse diffusion is used for text-to-speech generation in \cite{huang2022prodiff}. Leveraging DDIM sampling, a new student diffusion sampler learns at each iteration of the distillation process how to perform reverse diffusion using half as many steps as the current teacher. The resulting distilled sampler generates speech with similar quality as the original sampler using 64x more steps.

The noise variance schedule and time discretization used for reversed diffusion can also be optimized to reduce the number of steps, instead of being pre-defined. In \cite{lam2022bddm}, the schedule is learned by training an auxiliary hyper-network on top of existing denoising diffusion models. The resulting approach enables impressive speech generation results in as few as three reverse diffusion steps. 

Finally, some auxiliary losses and training schemes are designed to ensure that the diffusion states remain as close as possible to the domain seen by the score network during training, thereby mitigating the so-called drifting bias. Lay et al. \cite{lay2023single} propose a two-stage training method for diffusion-based speech enhancement following such a concept. The score network is first trained with denoising score matching and then fine-tuned to overfit a particular reverse diffusion sampler, by matching the final estimate of the solver to the clean speech target. High-quality speech enhancement is obtained with as few as one reverse diffusion step, reaching real-time computational complexity.

\textit{Causal processing}: In real-time acoustic communications (e.g., hearing aids), future information can not be used to process the current signal which means processing must be causal. Diffusion models can be adapted for causal processing, as in~\cite{richter2024causal}, where the convolutional score network architecture and the audio level normalization procedure are modified to meet causality requirements.

\subsection*{Robustness to adverse conditions}

Artifacts produced by diffusion models can differ in nature from those produced by statistical signal processing methods or predictive deep learning models. It was observed in \cite{richter2023speech} that speech enhancement diffusion models tend to hallucinate for negative input \acp{snr}, i.e., when noise dominates clean speech. This can lead to speech inpainting in noise-only regions, breathing and gasping artifacts, or the introduction of phonetic confusion, which may have a negative impact in real-world applications. This behavior can be mitigated by introducing external modalities such as video in~\cite{richter2023audio}, where lip movements are analyzed to determine the phoneme used as conditioning for score estimation guidance. Alternatively, as presented in StoRM \cite{Lemercier2022storm} the input \ac{snr} can be first increased by using a predictive deep learning model to remove parts of the noise, at the potential cost of speech distortions. A generative diffusion model is then used to reconstruct the noisy and distorted speech, which was shown to help avoid hallucination effects and thus increase the robustness to challenging conditions.

Generative pre-training is another approach to increase robustness to outliers. It involves using a pretext task such as masked modeling to train the diffusion model in a self-supervised fashion. Masked modeling involves randomly masking some regions of audio and instructing the model to fill in those masked sections using the available context information, i.e., the non-masked regions. This pre-trained model can then be fine-tuned for a particular downstream task (e.g., speech enhancement, music restoration, etc.) using a supervised setting. Liu et al. \cite{liu2024generative} show that their diffusion model SpeechFlow benefits from generative pre-training, as it increases its robustness to adverse scenarios such as noise-dominated utterances in speech enhancement. They also notice that generative pre-training consistently increases performance for most speech restoration tasks. 

Finally, running several realizations of the reverse diffusion process and measuring the empirical standard deviation of the obtained estimates can provide the user with a natural measure of uncertainty, which can help detect outliers and estimate the robustness of the approach on the given task.

\section*{Conclusion}

This article discussed diffusion models as deep conditional generative models for audio restoration. We suggested that diffusion models can be considered as serious candidates for model-based audio processing, as we recalled that domain knowledge can be injected into various aspects of their design such as parameterization of diffusion trajectories, or modeling of a measurement likelihood for posterior sampling with diffusion priors. By categorizing the various forms of conditioning proposed in diffusion approaches---namely input conditioning, task-adapted processes, and external conditioning---we highlight the structural flexibility of diffusion models and their resulting appreciable degree of interpretation. In particular, looking at audio restoration under the scope of solving inverse problems, we showed that we can combine diffusion models with Bayesian tools and stochastic optimization, thereby leveraging various parameterizations of degradation operators for informed and blind inverse problems. The quality of diffusion-based audio generation is remarkable, and although this can be originally outbalanced by disadvantages regarding practical requirements, e.g., robustness to adverse conditions or inference speed, we exposed several approaches and studies solving these drawbacks. We believe these solutions can be combined to yield robust, fast diffusion models for real-time acoustic communications. 

\section*{Acknowledgments}

This work has been funded by the German Research Foundation (DFG) in the transregio project Crossmodal Learning (TRR 169), DASHH (Data Science in Hamburg - HELMHOLTZ Graduate School for the Structure of Matter) with the Grant-No. HIDSS-0002, and NordicSMC (Nordic Sound and Music Computing Network) with NordForsk project 86892. 

\section*{Authors}

\noindent \textbf{Jean-Marie Lemercier}
(jeanmarie.lemercier@uni-hamburg.de) received an M.Eng in Electrical Engineering in 2019 from Ecole Polytechnique, Paris, France. In 2020, he received a M.Sc. in Communications and Signal Processing from Imperial College London, London, United Kingdom. He is currently a PhD student in the Signal Processing group at Universität Hamburg, 20148 Hamburg, Germany. For his work on speech enhancement with diffusion-based generative models, he was the recipient of the Association for Electrical, Electronic, and Information Technologies Informationstechnische Gesellschaft (VDE ITG) Award 2024. He received the Best Student Paper Award at IWAENC 2024. His research interests span machine learning-based speech enhancement and dereverberation for hearing devices applications. Recent works also include the design and analysis of diffusion-based generative models for various speech restoration tasks. He is a Student Member of IEEE.

\vspace{1em}

\noindent \textbf{Julius Richter} (julius.richter@uni-hamburg.de) received a B.Sc. and M.Sc. in Electrical Engineering in 2017 and 2019 from the Technical University of Berlin, Germany. He is currently a PhD student in the Signal Processing group at Universität Hamburg, 20148 Hamburg, Germany. For his work on speech enhancement with diffusion-based generative models, he was the recipient of the Association for Electrical, Electronic, and Information Technologies Informationstechnische Gesellschaft (VDE ITG) Award 2024. His research interests include deep generative models and multi-modal learning with applications to audio-visual speech processing. He is a Student Member of IEEE.

\vspace{1em}

\noindent \textbf{Simon Welker}
(simon.welker@uni-hamburg.de) received a B.Sc. in Computing in Science (2019) and an M.Sc. in Bioinformatics (2021) from Universität Hamburg, Germany. He is currently a Ph.D. student in the Signal Processing Group, Universität Hamburg, 20148 Hamburg, Germany, and the Coherent Imaging Division, Center for Free-Electron Laser Science, Deutsches Elektronen-Synchrotron DESY, 22607 Hamburg, Germany. For his work on speech enhancement with diffusion-based generative models, he was the recipient of the Association for Electrical, Electronic, and Information Technologies Informationstechnische Gesellschaft (VDE ITG) Award 2024.
His research interests include machine learning techniques for
solving inverse problems that arise in speech processing and
X-ray imaging. He is a Student Member of IEEE.

\vspace{1em}

\noindent \textbf{Eloi Moliner} (eloi.moliner@aalto.fi) received his B.Sc. degree in telecommunications technologies and services engineering in 2018 and his M.Sc. degree in telecommunications engineering in 2021 from the Polytechnic University of Catalonia, Spain. He is currently a Ph.D. candidate at the Acoustics Lab of Aalto University, 02150 Espoo, Finland. He is the winner of the Best Student Paper Award of the 2023 IEEE ICASSP conference. He received the Best Student Paper Award at IWAENC 2024. His research interests include digital audio restoration and audio applications of machine learning.

\vspace{1em}

\noindent \textbf{Vesa Välimäki} (vesa.valimaki@aalto.fi) received his D.Sc. degree in electrical engineering from the Helsinki University of Technology in 1995. In 1996, he was a post-doctoral research fellow at the University of Westminster, London, U.K. In 2008–2009 he was a visiting scholar at Stanford University. He currently is a full professor of audio signal processing and vice dean for Research at Aalto University, 02150 Espoo, Finland. He is a fellow of the Audio Engineering Society and of the Asia-Pacific Artificial Intelligence Association, and is the editor-in-chief of the \emph{Journal of the Audio Engineering Society}. His research interests include the application of machine learning and signal processing to audio technology. He is a Fellow of IEEE.

\vspace{1em}

\noindent \textbf{Timo Gerkmann} (timo.gerkmann@uni-hamburg.de) received his Dr.-Ing. degree in electrical engineering and information sciences from Ruhr-Universität Bochum, Germany. He is a professor for signal processing with the Universität Hamburg, 20148 Hamburg, Germany. He was the recipient of the Association for Electrical, Electronic, and Information Technologies Informationstechnische Gesellschaft (VDE ITG) Award 2022. He served in the IEEE Signal Processing Society Technical Committee on Audio and Acoustic Signal Processing and is currently a senior area editor of the \emph{IEEE/ACM Transactions on Audio, Speech and Language Processing}. His research interests include statistical signal processing and machine learning for speech and audio applied to communication devices, hearing instruments, audio-visual media, and human-machine interfaces. He is a Senior Member of IEEE.

\bibliographystyle{ieeetr}
\bibliography{biblio}

\end{document}